\documentclass[aps, prb, twocolumn, groupedaddress, showpacs, showkeys,
longbibliography ]{revtex4-1}

\usepackage{graphicx}
\usepackage{hyperref}
\usepackage{longtable}
\usepackage{color}

\newcommand\eqref[1]{Eq. (\ref{#1})}
\newcommand\figref[1]{Fig. \ref{#1}}
\newcommand\tabref[1]{Tab. \ref{#1}}

\begin{document}


\title{Charge Localization and Ordering in A$_2$Mn$_8$O$_{16}$ Hollandite Group Oxides:
Impact of Density Functional Theory Approaches}


\author{Merzuk Kaltak}
\email[]{merzuk.kaltak@stonybrook.edu}
\author{Marivi Fern{\'a}ndez-Serra}
\email[]{maria.fernandez-serra@stonybrook.edu}
\affiliation{Department of Physics and Astronomy, SUNY Stony Brook University,
New York 11974-3800, USA}
\affiliation{Institute for Advanced Computational Science, SUNY Stony Brook
University, Stony Brook, New York 11794-3800, USA} 
\author{Mark S. Hybertsen}
\email[]{mhyberts@bnl.gov}
\affiliation{Center for Functional Nanomaterials, Brookhaven National
Laboratory, Upton, New York 11973-5000, USA} 


\date{\today}

\begin{abstract}
The phases of A$_2$Mn$_8$O$_{16}$ hollandite group oxides
emerge from the competition between ionic interactions, Jahn-Teller effects,
charge ordering, and magnetic interactions. 
Their balanced treatment with feasible computational approaches
can be challenging for commonly used approximations in Density Functional Theory.
Three examples (A = Ag, Li and K) are studied
with a sequence of different approximate exchange-correlation
functionals. Starting from a generalized gradient approximation (GGA),
an extension to include van der Waals interactions and
a recently proposed meta-GGA are considered.
Then local Coulomb interactions for the Mn $3d$ electrons are more explicitly considered
with the DFT+$U$ approach. Finally selected results from a hybrid functional approach provide a reference.
Results for the binding energy of the A species in the parent oxide
highlight the role of van der Waals interactions.
Relatively accurate results for insertion energies can be achieved with a low $U$ and a high $U$ approach.
In the low $U$ case, the materials are described as band metals
with a high symmetry, tetragonal crystal structure.
In the high $U$ case, the electrons donated by A result in formation of
local Mn$^{3+}$ centers and corresponding Jahn-Teller distortions
characterized by a local order parameter.
The resulting degree of monoclinic distortion depends on charge ordering
and magnetic interactions in the phase formed.
The reference hybrid functional results show charge localization and ordering.
Comparison to low temperature experiments of related compounds suggests
that charge localization is the physically correct result for the hollandite group oxides studied here.
Finally, while competing effects in the local magnetic coupling are subtle,  
the fully anisotropic implementation of DFT+$U$ gives the best overall agreement
with results from the hybrid functional. 
\end{abstract}

\maketitle

\section{Introduction}\label{sec:Introduction}
\begin{figure}
\includegraphics{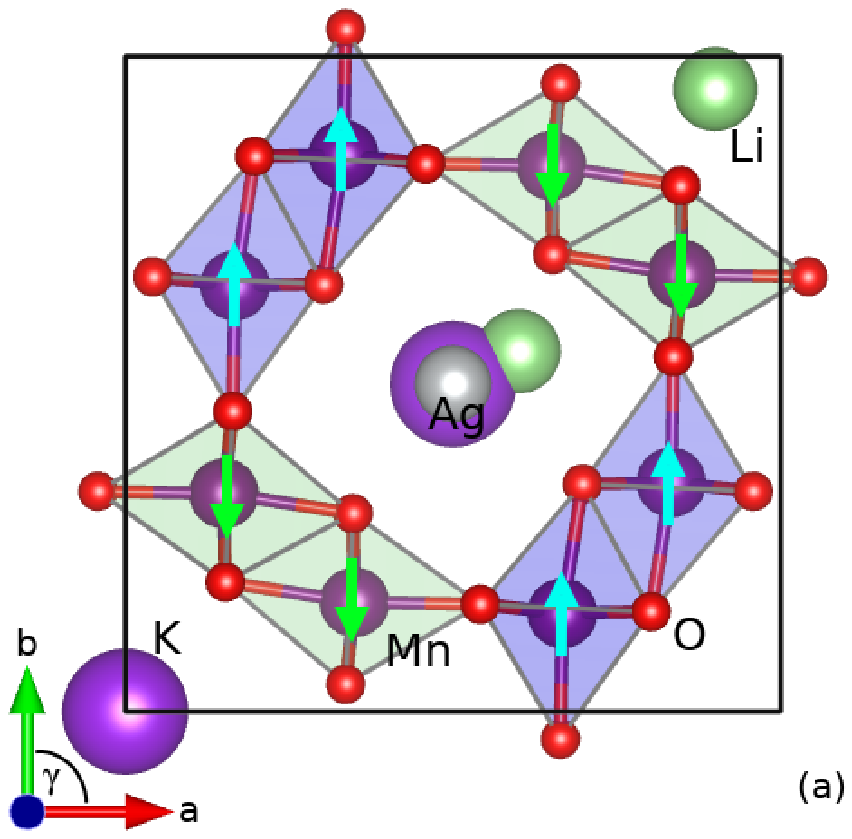}
\includegraphics{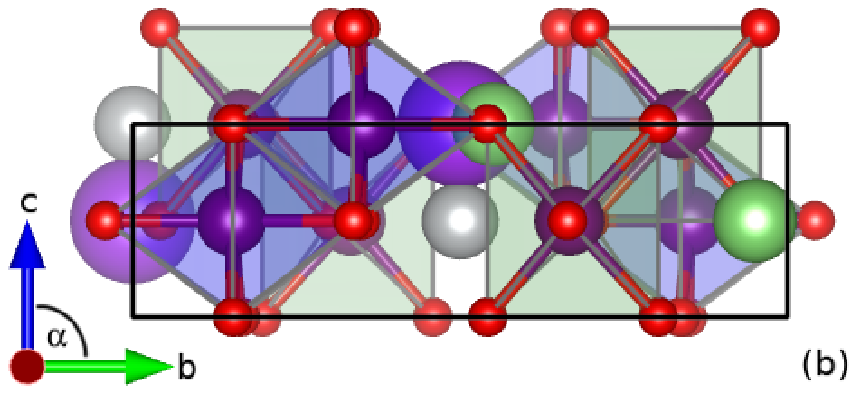}
\caption{\label{fig:UnitCell} (color online)  $I 4/m$ unit cell of
Mn$_8$O$_{16}$ with Wyckoff positions of Ag ($2a$), K ($2b$) and Li ($8h$)
visualized with VESTA\cite{VESTA} in two views, (a) and (b). Here the semitransparent
blue and green shaded MnO$_6$ octahedra represent opposite spin orientation 
for the specific, antiferromagnetic order found in this work.  
Other choices for magnetic order are
discussed further in the text.} 
\end{figure}

Manganese oxide minerals form in structures with an astonishing natural
diversity and have a variety of practical applications.\cite{Post99} The
hollandite group includes such specific minerals as hollandite,
Ba$_x$Mn$_8$O$_{16}$, and cryptomelane, K$_x$Mn$_8$O$_{16}$.  The backbone
structure consists of edge-sharing MnO$_6$ octahedra that form double chains.
These, in turn, connect through corner shared oxygens to form tunnels, as
visualized in \figref{fig:UnitCell}, resulting in an inherently one-dimensional
structural feature. The additional ions such as Ba$^{2+}$, Ag$^+$, K$^+$,
Na$^+$, Li$^+$, {\em etc.}, occupy the larger cross-section tunnels.  Water may
also be incorporated in the tunnels. The ideal form of hollandite incorporates
up to two ions per formula unit $(x=2)$ and the space group of the tetragonal
cell is $I4/m$.\cite{Bystrom50} The $\alpha-$MnO$_2$ structure corresponds to
$x=0$.

More broadly, the hollandite structure forms with tetravalent metal ions at
the octahedral centers, {\em e.g.}, with Mn$^{4+}$, Ti$^{4+}$, Cr$^{4+}$ and
V$^{4+}$.  With the incorporation of ions in the channels, there must be a
corresponding reduction of ions in the backbone as well.  The result can be a
mix of Mn$^{4+}$ and Mn$^{3+}$ sites in the backbone or the inclusion of
trivalent metal ions explicitly during synthesis, for instance
Fe$^{3+}$.\cite{Biagioni13} Naturally occurring minerals in the hollandite group
often involve complex mixtures of cations both in the tunnels and within the
octahedra. Depending on the ratio of the average ionic radii in the two
positions, the crystal symmetry is observed to be lowered to monoclinic
($I2/m$), with the general trend indicating that relatively smaller ions in the
tunnels correspond to monoclinic structures.\cite{Post82}

The hollandite group manganese oxides, and closely related structures with
different sized tunnels, represent a tunable system with nanoscale,
one-dimensional pores.\cite{Brock98} The one-dimensional pores present a
template for both fundamental studies of the impact of dimensionality on
processes such as ionic diffusion\cite{Beyeler76,Bernasconi79} and an internal
surface support for chemical processes.  In particular, hollandite group
manganese oxides have been studied for use in catalysis\cite{Chen07,
Dharmarathna12} and extensively for potential utility as cathodes in batteries
based on Li$^+$, Na$^+$, and Mg$^{2+}$.  \cite{ROSSOUW1992, Johnson97a, Dai00,
BARBATO2001, Kijima2005, Johnson2007, Zhang12, Ling12, Trahey13, Islam13, Yuan15, Yuan16, Xu17,
Yang17,Doeff94, Sauvage07, Kim12, PerezFlores14, Huang17, Zhang2012, Huang16} 
Facile synthesis of Ag-hollandite under mild conditions\cite{Chen07,Zhu2010} has made
it more readily available for electrochemical studies.  Furthermore, control of
Ag insertion during synthesis results in clear changes in sample composition and
structure with correlated changes in electrochemical
characteristics.\cite{Huang17,Takeuchi12,Takeuchi13} In particular, there is an
interplay between Ag content, oxygen vacancy concentration and the diameter of
the nanorod morphology crystallites that are formed with the long axis parallel
to the tunnel direction.\cite{Wu15} Understanding the relationship of these
structural degrees of freedom and the electrochemical response remains as an
on-going challenge.  The present work is motivated by our need to assess
methods based on Density Functional Theory (DFT)
for use in the broader exploration of phases formed in the
Li$_x$Ag$_y$Mn$_8$O$_{16-z}$:(H$_2$O)$_w$ family of hollandite-derived
materials.

The fundamental electronic structure of hollandite structure oxides with the
chemical formula A$_x$M$_8$O$_{16}$ can be quite subtle, depending on the way
the reduction implied by the tunnel cation A is accommodated.  Formally, the
backbone transition metal M is in a mixed valence state between M$^{4+}$ and
M$^{3+}$.  Furthermore, according to Hund's rule, ions such as Mn$^{4+}$ and
Mn$^{3+}$ with three and four $d$ electrons respectively should be in a high spin
state locally.  The octahedra in the parent $\alpha-$MnO$_2$ structure ($I4/m$)
are close to ideal, with the six Mn-O bond lengths varying minimally (range of
0.02 \AA~ near a value of 1.90 \AA),\cite{Kijima2004} 
implying a filled majority spin $t_{2g}$ manifold. The fourth $d$ electron
for a local Mn$^{3+}$ will go to the $e_g$ manifold setting up the possibility
for a local Jahn-Teller distortion to lower the electronic energy.  
However, the degree to which the excess electrons donated by the tunnel cations
localize and local structural distortions reduce the symmetry, must emerge
from competing factors: ionic radius ratio, local Coulomb interactions, the Jahn-Teller
effect, and magnetic interactions.
These factors will vary with the backbone metal cation M and the tunnel cation A.

Not surprisingly, characterization of specific hollandite structure oxides has
revealed phase diagrams with distinct magnetic and conducting phases.  Two cases
have been closely studied as a function of temperature.  Near stoichiometric
K$_2$V$_8$O$_{16}$ has a first order transition to a monoclinic phase at 170 K
understood in terms of charge ordering of localized V$^{3+}$, dimerization along
the tunnel axis and strong spin interactions.\cite{Isobe06, Komarek11, Kim16,
Isobe09} In  K$_2$Cr$_8$O$_{16}$, an unusual sequence showed ferromagnetic
ordering at 180 K followed by a metal-insulator transition at 95
K.\cite{Hasegawa09} The low temperature phase has a subtle monoclinic
distortion, but there is no evidence for charge localization.\cite{Toriyama11, Kim14} For
hollandite group manganese oxides, the magnetic measurements at low temperature
have not been so clear-cut, with the latest data interpreted to indicate a
spin-glass phase.\cite{Yamamoto74, Strobel84, Sato99, Ishiwata06, Luo10,
Barudzija16} While naturally occurring cryptomelane exhibited a mild monoclinic
distortion,\cite{Post82} room temperature characterization of synthesized
hollandite group oxides showed a tetragonal phase ($I4/m$) for K, Li  and Ag
cases.\cite{Vicat1986, Barudzija16, Kadoma2007, Jansen1984} In the Ba case, a
monoclinic phase ($I2/m$) was found.\cite{Ishiwata06} However, no detailed
studies at low temperature have been performed to our knowledge.  Several
studies based on DFT approaches have considered the
ground state phase for $\alpha-$MnO$_2$ and ion insertion,
also with some disagreement in the results regarding monoclinic distortion and
magnetic order.\cite{Ling12, Islam13, COCKAYNE2012, Crespo13a, Ochoa16,
Kitchaev17} Some of the differences can likely be traced to different choices in
the approximate exchange-correlation functional.

In this paper, we revisit the application of the DFT-based tool-box to this
challenging problem by critically comparing results obtained from several
approximations to the treatment of exchange and correlation. We focus on
$\alpha-$MnO$_2$ and the insertion of K, Ag and Li to form A$_2$Mn$_8$O$_{16}$.
While in many cases synthesis of the ideal stoichiometry with two tunnel cations
per unit cell has been difficult, we study this case to eliminate the additional
complication of sampling configurations for occupancy of the A atom positions,
as would be necessary for lower concentrations.  

Starting with the version of the generalized gradient approximation (GGA)
developed by Perdew, Burke and Ernzerhof (PBE),\cite{Perdew96} 
we consider the impact of van der Waals (vdW)
interactions through the Opt-B88 functional.\cite{OptB88} 
With these as a base,
we then study the inclusion of the Hubbard correction term
for the Mn $3d$ electrons, using the DFT+$U$ approach,\cite{Anisimov91, Anisimov97} 
characterizing the 
physical and electronic structure as a function of
the value of $U$ as well as the impact of the spherical approximation.\cite{Dudarev98}  
This method has proven to be a cost effective way to
substantially improve the treatment of systems with strong local Coulomb
interactions and Jahn-Teller effects.\cite{Anisimov97,Liechtenstein95} For
comparison, we also consider the Heyd, Scuseria and Ernzerhof (HSE) hybrid
functional approach.\cite{HSE03,HSE06} Previous studies of perovskites suggest that it
handles the subtleties of the open shell transition metals relatively
accurately, albeit with increased computational cost.\cite{Franchini14} Finally,
we consider the recently developed strongly constrained and appropriately normed
(SCAN) meta-GGA for the exchange-correlation
functional.\cite{SCAN} 
Analysis of DFT+$U$ and HSE for manganese oxide phases specifically
suggest that both can account for structure and magnetic phases.\cite{Franchini07,Tompsett12,Lim16}
Results for SCAN show that it may be an accurate and efficient alternative for
manganese oxide energetics.\cite{Kitchaev16} 
Our study focuses on the treatment of physical properties
resulting from mixed valence between Mn$^{4+}$ and
Mn$^{3+}$.

Across the methodological choices, we examine energetics, charge localization,
charge ordering, structural distortion and magnetic order.  Using results based
on HSE as a reference, we characterize the accuracy of the other functionals for
the addition energy associated with Ag, Li and K, fundamental to predicting
electrochemical properties.  We then explore the implications for the crystal
structure, finding a critical role for charge localization to form Mn$^{3+}$ centers.  
Charge localization is predicted by the HSE calculations 
and the GGA+$U$ methods for moderate to large values of $U$.  Correspondingly,
electronically driven Jahn-Teller distortions also emerge, 
characterized by a local order parameter.
They result in strong
monoclinic distortions of the unit cell.  We discuss the role of charge ordering
in this case and the implications for the observed crystal structures.  We also
briefly discuss the magnetic phases formed.  In agreement with prior work and
experiment, we find that the competition between magnetic coupling mechanisms is
delicate and presents a challenge.
We do find that including the anisotropy of the Coulomb interaction
in DFT+$U$ with larger values of $U$ 
gives results closer to those from the HSE hybrid functional.

This paper is organized as follows. In the following section we summarize the
expected Jahn-Teller effect in the hollandite group oxides,
introduce a local order parameter to measure
the corresponding distortion of the tetragonal unit cell and give a detailed
description of all the DFT methods employed. In Sect. \ref{sec:Results} we
discuss our results comparing the various different DFT approaches for
energetics, crystal structure, charge localization and charge ordering. We then
discuss these results in Sect. \ref{sec:Discussion} in light of the literature
for related compounds and compare two reasonable choices of method to study more
general phase formation in Li$_x$Ag$_y$Mn$_8$O$_{16-z}$:(H$_2$O)$_w$. In the
last section we summarize our work.

\section{Theory and Method}\label{sec:Theory}
\subsection{Structure and Jahn-Teller Effect}\label{sec:JahnTeller}
The basic unit cell depicted in \figref{fig:UnitCell}, which represents the
formula unit A$_2$Mn$_8$O$_{16}$, will be the building block for our studies.
The edge-sharing octahedra form the walls of the large $2\times2$ tunnels and
point-sharing oxygen ions are the corners of the smaller $1\times1$ tunnels.
Typically the $1\times1$ tunnels are too narrow to host large concentrations of
guest ions,\cite{Yuan15} so that the inserted Li ions, or the structurally
supporting ones like Ag or K, are located inside the larger $2\times2$ tunnels.
Previous DFT studies on Li$_x$Mn$_8$O$_{16}$\cite{Islam13, Ling12} have shown
that Li prefers to occupy the off-center positions at the concentration $x=2$.
Ag and K prefer a more symmetrical coordination, but generally
with a different position along the tunnel direction.
These are illustrated in
\figref{fig:UnitCell}.\cite{Ochoa16, LI2013120, Yuan15} 

\begin{figure}
\includegraphics{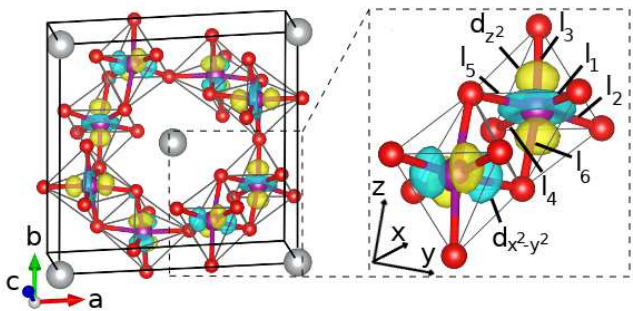}
\caption{\label{fig:LocalWannier}
(color online) Unit cell (left) and zoom on a pair of edge-sharing octahedra (right)
with illustrations of the local Wannier orbitals for the Mn $e_g$ states.
Local coordinates are defined for reference, along with bond
length definitions for \eqref{eq:Index}.}
\end{figure}

In the absence of any cations in the tunnels, the structure of hollandite is
tetragonal and has $I 4/m$ symmetry (space group 87). In this case the cell
contains exclusively Mn$^{4+}$ ions with an almost uniform octahedral crystal
field. This field splits the degenerate atomic $3d$ states of Mn into the
three-fold degenerate, occupied $t_{2g}$ states $d_{xy}, d_{xz}$ and $d_{yz}$ and the
double degenerate, unoccupied $e_{g}$ states $d_{z^2}$ and $d_{x^2-y^2}$.
Examples of the $e_g$ states in the local framework of the octahedra are
illustrated in \figref{fig:LocalWannier}.  In the ferromagnetic (FM) solution
the $t_{2g}$ electrons of all Mn ions are aligned in parallel. In the case of
anti-ferromagnetic (AF) order, there are a number of different ways for the spins to
be organized, both within the basic unit cell shown in \figref{fig:UnitCell}, as
well as simple generalizations, such as a doubling along the indicated $c$ axis.
Several alternatives have been enumerated in recent literature.\cite{Crespo13b}
In agreement with Ref. \onlinecite{Crespo13a}, we find that the ordering pattern
designated C2-AFM, where the spin sign alternates between corner-sharing Mn
ions, as illustrated in \figref{fig:UnitCell}, has lower energy than the
competing patterns. Results designated as AF in this paper refer to this order.  

Independent of magnetic ordering, the atoms A in the tunnel typically ionize
and transfer electrons
to the Mn$^{4+}$ cations in the oxide backbone.  One possibility is that the
extra electrons partially occupy a sufficiently dispersive band, or multiple
bands, such that they are delocalized.  Another possibility is that the excess
electrons are localized, in the extreme limit each one resulting in a local
Mn$^{3+}$ cation.  In the present case for monovalent atoms A in the tunnel, this
can formally be written (A$^+$)$_2$(Mn$^{3+}$)$_2$(Mn$^{4+}$)$_6$(O$^{2-}$)$_{16}$.  In a
solution with local Mn$^{3+}$ cations, the ordering must also be determined.
Several possible examples of ordering within the basic hollandite unit cell and
small supercells ($\sqrt{2} \times \sqrt{2} \times 1$ and $1 \times 1 \times 2$)
are discussed in detail in \ref{sec:ChargeOrdering}.

The $3d^4$ Mn$^{3+}$ cation should be high spin according to Hund's rule so the
extra electron will occupy one of the nearly degenerate $e_g$ states.  Following
the usual Jahn-Teller argument,\cite{JahnTeller1937}  the local octahedron can
distort to a lower symmetry solution, split the degenerate $e_g$ states through
a distorted crystal field and gain energy by singly occupying the lower energy
state.  The net energy gain is limited by the local elastic energy cost of the
distortion.  Two orthogonal distortions, $Q_2$ and $Q_3$, were introduced by Van
Vleck to span distinct symmetry reduction pathways from the ideal octahedron
that couple to the $e_g$ states.\cite{VanVleck1939} Generalized to the slightly
distorted initial octahedra found in $\alpha-$MnO$_2$ and written in terms of
the Mn-O bond lengths defined in \figref{fig:LocalWannier}, these are
\begin{eqnarray}\label{eq:Index}
Q_2 &=& \frac{1}{\sqrt2}\left(l_1+l_4-l_2-l_5\right) \\
Q_3 &=& \frac{1}{\sqrt6}\left( 2l_3+2l_6 -l_1-l_4-l_2-l_5\right).
\end{eqnarray}
Here $l_{3,6}$ are the MnO bond lengths in the locally defined $z$ direction and $l_{1,4},
(l_{2,5})$ the corresponding bond lengths in the $x(y)$
direction.\cite{VanVleck1939} The $Q_2$ order parameter measures the local
orthorhombic distortion, and $Q_3$ the tetragonal distortion, of an octahedral
crystal field.\cite{VanVleck1939}

Each solution with local Mn$^{3+}$ cations can be expected to have some degree
of local Jahn-Teller distortion that can be quantified through $Q_2$ and $Q_3$.
Furthermore, these distortions will couple to the crystal structure
cooperatively in a particular way determined by how they are ordered, resulting
in a final, possibly distorted unit cell.  Finding the ground state at zero
temperature requires sampling different patterns of charge order to determine
the one with lowest energy.  
Target charge order is probed by locally imposing the distortion
and relaxing the ionic positions to determine if the order is self consistently sustained.

\subsection{Technical Details}\label{sec:TechnicalDetails}
All DFT calculations have been performed with the Vienna \textit{Ab initio}
Simulation Package (VASP).\cite{Kresse99}  
Specific exchange-correlation functionals utilized include
PBE,\cite{Perdew96} Opt-B88,\cite{OptB88} and SCAN.\cite{SCAN} 
For the PBE and Opt-B88 calculations, the GGA+$U$ approach was used, specifically
applied to the $3d$ electrons of Mn. 
Survey calculations were done using the spherical
approximation.\cite{Dudarev98} In this case, one effective Coulomb interaction
$U_{\rm eff}=U-J$ is required, here treated as a parameter for analysis.
For comparison, selected PBE+$U$ calculations were performed retaining the anisotropy
of the Coulomb interaction, specifically with $U=6$ eV and $J=1$ eV
and to be compared to $U_{\rm eff}$=5.0 eV.
These values are similar to those chosen in a recent study of
magnetic order in $\beta-$MnO$_2$\cite{Tompsett12}
while the choice of $U$ is somewhat larger than those identified
in a recent study of five manganese oxides.\cite{Lim16}
For the balance of this paper, we will use compact notation
to distinguish the GGA+$U$ calculations according to the
choice of GGA, the parameter values and the use of the spherical approximation,
{\em e.g.}, Opt-B88+$U_{\rm eff}$=1.6 eV in the spherical approximation
and PBE+($U$=6, $J$=1 eV) considering full anisotropy.

For the projector augmented wave\cite{Blochl94} basis set, 
a cutoff of $520$ eV was employed in combination
with the semi-core potentials Mn:$3s^23p^64s^23d^5$, O:$2s^22p^4$,
Ag:$4s^24p^65s^14d^{10}$, K:$3s^23p^64s^1$ and Li:1s$^2$2s$^1$ resulting in unit
cells with roughly 250 electrons. The Brillouin zone of the unit cell was
sampled with a $\Gamma$-centered, $2\times 2\times6$ 
grid of $k$ points. The Methfessel-Paxton smearing
method of order 1 was used with a smearing parameter of $\sigma$=0.1 eV.
Self-consistency was converged to a total energy convergence
criterion of $10^{-6}$ eV. Using these settings we have relaxed all
structures with the conjugate gradient algorithm until the residual force
(acting on each individual ion) was less than $20$ meV/\AA. 
As needed, cell shape was also relaxed, with accuracy in this procedure
assured by the large basis set and by checking for changes upon restart
following initial convergence of the cell parameters.
After performing
convergence tests with respect to $k$ point sampling and energy cutoff, we estimate
that PBE, Opt-B88 and SCAN energy differences reported below are given with a
precision of 5 meV/cell or smaller. 

For the hybrid functional calculations, we have chosen HSE in the formulation of
2006 with 25\% Hartree-Fock exchange and a range separation parameter of
$\omega=0.2$ \AA$^{-1}$.\cite{HSE06} Here the relaxations have been done
employing a PAW basis set cutoff of $450$ eV in combination with a
$1\times1\times4$ $k$ point sampling and a residual force threshold of $20$
meV/\AA. These cutoffs were slightly reduced, for computational efficiency,
relative to the settings used otherwise. Note that semi-core electrons are
retained, following experience that shows the importance of core-valence
exchange, particularly for $d$ electron cases.\cite{Marini01, Paier06, Engel09}
For the relaxed unit cells, total energies were recalculated with the same
settings as those used for the PBE, Opt-B88 and SCAN calculations. Only small
deviations in the forces and in the binding energies of A in A$_2$Mn$_8$O$_{16}$
were observed. We estimate that HSE-based relative energies reported below are
given with a precision of 20 meV/cell or better.

The localized electronic structure was studied using a basis set of maximally
localized Wannier functions.\cite{Marzari97} 
The projection
of the PAW basis set on the valence states Mn:$d$, O:$p$, K:$s$ and Li:$s$
orbitals was done using the wannier90 tool with the VASP2WANNIER90
interface.\cite{wannier90} For Ag the inclusion of
the $d$ states in addition was necessary to reproduce the band structures with
the Wannier interpolation method. The projected density of states (PDOS) has
been interpolated using a $k$ point grid of $6\times6\times24$ points.    

\section{Results}\label{sec:Results}

\subsection{Binding energies}\label{sec:BindingEnergy}
\begin{figure}
\includegraphics{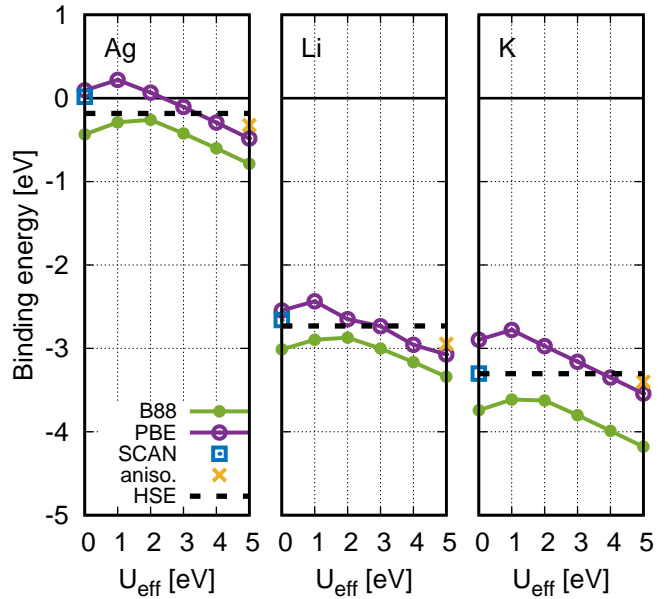}
\caption{\label{fig:BindingEnergy}
(color online) Binding energy $E(A)$ of A=Ag, K and Li, in A$_2$Mn$_8$O$_{16}$, as defined in
\eqref{eq:BindingEnergy}, based on Opt-B88+$U_{\rm eff}$ (points)
and PBE+$U_{\rm eff}$ (circles) as function of $U_{\rm eff}$. 
The other
binding energies shown are predicted by the HSE hybrid functional (dashed line),
SCAN (squares) and PBE+($U$=6, $J$=1 eV) (crosses).}
\end{figure}

We first consider the binding energies of A in A$_2$Mn$_8$O$_{16}$. For
this purpose we have calculated
\begin{equation}\label{eq:BindingEnergy}
E(A) = \frac{E_A - E_0 -2\mu_A}{2},
\end{equation}
where $E_A$ is the energy of the unit cell of A$_2$Mn$_8$O$_{16}$, $E_0$ the
energy of pristine hollandite and $\mu_A$ the corresponding chemical potential
of A. Here, both Mn$_8$O$_{16}$ and A$_2$Mn$_8$O$_{16}$ are computed for the AF
ordering shown in \figref{fig:UnitCell} and correspondingly with the lowest
energy charge ordering found.  For the A chemical potential, we have chosen the
energy per atom of bcc-Li, bcc-K and fcc-Ag metal, respectively.  Thus
\eqref{eq:BindingEnergy} measures the difference in total electronic energy
({\em i.e.} $T=0$ K and no correction for zero-point motion) of the reaction
Mn$_8$O$_{16}+2\times$A$\to$ A$_2$Mn$_8$O$_{16}$. That is, negative (positive)
energies $E(A)$ indicate a spontaneous formation (decomposition) of
A$_2$Mn$_8$O$_{16}$ at zero temperature.  A comparison of $E(A)$ for a set of
different DFT functionals and values of $U_{\rm eff}$ is given in
\figref{fig:BindingEnergy}.

It can be seen that for all investigated cations, the Opt-B88+$U_{\rm eff}$ (points) and HSE
functional (dashed lines) yield negative binding energies implying a spontaneous
formation of A$_2$Mn$_6$O$_{16}$ at zero temperature. In contrast, SCAN
(squares) as well as PBE+$U_{\rm eff}$ (circles) for $U_{\rm eff}<2.5$ eV predict a positive
formation energy for Ag$_2$Mn$_8$O$_{16}$ implying that additional energetic
cost would be necessary to form silver hollandite at zero temperature.  This
suggests that there is insufficient description of vdW  interactions in PBE and
SCAN, a part of the correlation energy taken into account explicitly by Opt-B88
and empirically at short range with HSE.  Evidently this is important for the
chemical bonding of a noble metal like Ag in the tunnel.  The vdW term in the
exchange-correlation functional of Opt-B88+$U_{\rm eff}$ shifts the binding energies of all
three cations almost uniformly by $\approx 0.5$ eV downwards compared to PBE+$U_{\rm eff}$, as
seen by comparing the closed and open circles in \figref{fig:BindingEnergy}.
The same figure also reveals that in both PBE+$U_{\rm eff}$ and Opt-B88+$U_{\rm eff}$, the binding energy
decreases monotonically for all investigated cations for $U_{\rm eff}>2$ eV and
approaches the HSE result at specific values.  Also, inclusion of anisotropy 
with PBE+($U$=6, $J$=1 eV) slightly improves the results compared to HSE.

In applications, such as comparison of energies 
for different compositions to build a phase diagram, 
one value of $U_{\rm eff}$ must be selected for consistency.
Here, \figref{fig:BindingEnergy} provides a
calibration, using the HSE results as a reference.  The best match with HSE
binding energies for all three cations is obtained for the Opt-B88+$U_{\rm eff}$ functional
with moderate values of $U_{\rm eff}\le2$ eV and for the anisotropic
PBE+($U$=6, $J$=1 eV) method. A somewhat less accurate match is found with PBE+$U_{\rm eff}$ for
values of $U_{\rm eff}$ in the range 3 to 5 eV.

Over all, K has the strongest binding energy in the hollandite tunnel
followed by Li and the weakly bound Ag.  This trend does track the ionization
energy, a first measure for the energy gain possible upon formation of the ion
in the tunnel.  For instance, K has one of the lowest ionization energies in
nature\cite{NIST_ASD} (4.34 eV). This is followed by Li with 5.39
eV,\cite{NIST_ASD} which results overall in weaker ionic binding for Li compared
to K. In contrast, Ag has an ionization energy of 7.58 eV\cite{NIST_ASD} and,
correspondingly shows relatively weak ionic binding.  We have examined the
binding of other ions in the hollandite tunnel.  The binding energy for Na
(-3.10 eV) is found to be close to that for Li, in accordance with the trend
above (the ionization energy of Na is 5.14 eV\cite{NIST_ASD}). However, this
entire trend is violated by second row earth alkali elements like Mg, which has
an ionization energy of 7.65 eV, but a binding energy (-2.90 eV) similar to Li.
A simple model for the binding energy must include other factors beyond
ionization energy.

\subsection{Structure and Monoclinic Distortion}\label{sec:Structure} 
Next, we focus on the structure of pristine and doped hollandite compounds and
compare how selected, different DFT methods affect the corresponding lattice
constants. The results are reported in \tabref{tab:LatticeConstants}.
They all correspond to the AF order shown in \figref{fig:UnitCell}, except for the
case of PBE+($U$=6, $J$=1 eV) applied to Li$_2$Mn$_8$O$_{16}$
where FM order was lowest in energy. 
 \begin{table}
 \caption{For selected hollandite compounds, lattice constants and
angle $\gamma$ between primitive lattice vectors $a$ and $b$,
as calculated for several functionals in DFT.}
 \begin{ruledtabular}\label{tab:LatticeConstants}
 \begin{tabular}{r c c c c}
   & a [\AA] & b [\AA] & c [\AA] & $\gamma$ [deg] \\
  \hline
\multicolumn{5}{c}{Opt-B88}\\
      Mn$_8$O$_{16}$ & 9.58 &  9.78 & 2.85 & 89.43 \\
Ag$_2$Mn$_8$O$_{16}$ & 9.66 &  9.66 & 2.84 & 90.00 \\
 K$_2$Mn$_8$O$_{16}$ & 9.70 &  9.70 & 2.86 & 90.00 \\
Li$_2$Mn$_8$O$_{16}$ & 9.25 & 10.11 & 2.82 & 91.58 \\
\hline
\multicolumn{5}{c}{Opt-B88+$U_{\rm eff}$=1.6}\\
      Mn$_8$O$_{16}$ & 9.57 &  9.79 & 2.85 & 90.47 \\
Ag$_2$Mn$_8$O$_{16}$ & 9.70 &  9.70 & 2.85 & 90.00 \\
 K$_2$Mn$_8$O$_{16}$ & 9.49 & 10.18 & 2.88 & 91.57 \\
Li$_2$Mn$_8$O$_{16}$ & 9.39 & 10.18 & 2.84 & 92.10 \\
\hline
\multicolumn{5}{c}{PBE+($U$=6, $J$=1 eV)}\\
      Mn$_8$O$_{16}$ & 9.80 &  9.80 & 2.90 & 90.00 \\
Ag$_2$Mn$_8$O$_{16}$ & 9.51 & 10.35 & 2.90 & 91.77 \\
 K$_2$Mn$_8$O$_{16}$ & 9.64 & 10.34 & 2.91 & 91.03 \\
Li$_2$Mn$_8$O$_{16}$ & 9.61 & 10.35 & 2.89 & 91.73 \\
\hline
\multicolumn{5}{c}{SCAN}\\
      Mn$_8$O$_{16}$ & 9.54 & 9.72 & 2.83 & 90.41 \\
Ag$_2$Mn$_8$O$_{16}$ & 9.59 & 9.59 & 2.82 & 90.00 \\
 K$_2$Mn$_8$O$_{16}$ & 9.65 & 9.65 & 2.84 & 90.00 \\
Li$_2$Mn$_8$O$_{16}$ & 9.30 & 10.11& 2.81 & 91.88 \\
\hline
\multicolumn{5}{c}{HSE}\\
      Mn$_8$O$_{16}$ & 9.59 &  9.59 & 2.83 & 90.00 \\
Ag$_2$Mn$_8$O$_{16}$ & 9.41 & 10.11 & 2.84 & 91.53 \\
 K$_2$Mn$_8$O$_{16}$ & 9.39 & 10.16 & 2.85 & 91.49 \\
Li$_2$Mn$_8$O$_{16}$ & 9.40 & 10.02 & 2.81 & 92.19 \\
\hline
\multicolumn{5}{c}{Experiment}\\
      Mn$_8$O$_{16}^a$ & 9.777(2) &  & 2.8548(5) & 90.00 \\
Ag$_{1.8}$Mn$_8$O$_{16}^b$ & 9.725(7) &  & 2.885(2) & 90.00 \\
 K$_{1.33}$Mn$_8$O$_{16}^c$ & 9.866(3) &  & 2.872(1) & 90.00 \\
 Li$_x$K$_y$Mn$_8$O$_{16}^d$ & 9.81-9.89 &  & 2.855(2) & 90.00 \\
 \end{tabular}
 \end{ruledtabular}
$^a$Ref. \onlinecite{Kijima2004},
$^b$Ref. \onlinecite{Jansen1984},
$^c$Ref. \onlinecite{Vicat1986},
$^d$Ref. \onlinecite{Kadoma2007} 
 \end{table}

Starting with the results for the parent Mn$_8$O$_{16}$ structure, Opt-B88, SCAN
and Opt-B88+$U_{\rm eff}$=1.6 eV, yield a stable, slightly distorted structure
as lowest in energy. A tetragonal structure is also stable ($a=9.68$, 9.61 and
9.70 \AA~ respectively), but slightly higher in energy (1, 13 and 2 meV per cell
respectively). On the other hand, using PBE+($U$=6, $J$=1 eV) and HSE a tetragonal
structure has the lowest energy. Correspondingly, in those cases, a distorted structure can
also be stabilized, but it has a slightly higher energy per cell
(1 and 6 meV respectively). These are very small energy differences,
comparable to our estimated precision, but the relative ordering is robust upon increasing the k-point sample. 
These results are indicative of quite soft
degrees of freedom in the potential landscape that describes the hollandite
materials, particularly deformations related to bond angles centered on the
corner shared oxygen atoms.\cite{Crespo13a}
 
Upon introducing cations into the tunnel, Opt-B88 and SCAN yield almost
tetragonal structures for Ag and K, but predict a significant monoclinic distortion for Li. 
The latter distortion can be
explained by the ionic radius of Li and its off-center coordination within the
$2\times2$ tunnel walls.  In this location (see \figref{fig:UnitCell}) the pair
of Li ions reduce four Mn$^{4+}$ ions partially to Mn$^{3.5+}$.  Coulomb
interactions with these local Li ions act to distort the unit cell, above and
beyond Jahn-Teller effects.  This ion displacement driven 
distortion is absent for the larger cations Ag
and K, located in the tunnel center.
Introduction of Ag and K reduce the Mn$^{4+}$ ions uniformly to
Mn$^{3.75+}$. In the absence of any Hubbard term, these compounds behave like
simple band metals, with several empty bands of the parent Mn$_8$O$_{16}$
partially occupied.

As one increases the interaction $U_{\rm eff}$ in the GGA+$U_{\rm eff}$ methods, 
the details of the distorted structure induced by Li insertion change, but no
qualitative changes in the monoclinic cell emerge.  For Ag and K insertion,
there is a transition to a monoclinic cell at relatively small $U_{\rm eff}$.
For Opt-B88+$U_{\rm eff}$=1.6 eV reported in \tabref{tab:LatticeConstants}, the
K hollandite is distorted, whereas Ag hollandite remains tetragonal with
slightly larger lattice constants.  Increasing the interaction further triggers
a monoclinic distortion for Ag$_2$Mn$_8$O$_{16}$ as well.  For the large $U_{\rm eff}$
limit, with PBE+$U_{\rm eff}$, the Li, K and Ag hollandites are predicted to have very
similar monoclinic cells. For comparison, the hybrid HSE functional also
predicts monoclinic cells for all three ions, with $b-a$ values ranging from
0.62 \AA~for Li to 0.77 \AA~for K. 

In contrast, the measured crystal structures indicate tetragonal cells in all
cases.  None of the measured crystals had ideal cation concentration, but
variations in the lattice parameters among different cases is relatively small.
For example, X-ray measurements for Ag$_{1.22}$ and Ag$_{1.66}$ yielded $a =
9.770(2)$ and 9.738(2) \AA~respectively.\cite{Wu15} There was variation in the
stoichiometry of the compounds 
resulting from the ion exchange experiments reported in Ref.
\onlinecite{Kadoma2007}. The results were also affected by possible reactions with residual
water in the tunnels.  The range of lattice parameters reported reported for Li hollandite is
indicated in \tabref{tab:LatticeConstants}.  Trends with cation are on a small
scale, smaller than differences among calculations with different
exchange-correlation functionals.  However, the trend of reduced in-plane
lattice parameter $a$ for the Ag case, and the increase for K, are reproduced by
both Opt-B88 and SCAN functionals.  These are room temperature experiments.
Further discussion appears in Sect. \ref{sec:Discussion} below.

These results for K and Ag suggest that the structural deformation observed
in the calculations is electronically driven.  This may be significant for the
Li case as well.  The donated cation $s$ electrons are becoming more localized
with increasing $U_{\rm eff}$ value, forming two distinct Mn$^{3+}$ ions per unit
cell, so that at some point the system gains enough energy from orbital
reordering to distort the tetragonal symmetry. This cooperative Jahn-Teller
effect depends on the cation size as well as the coordination with the
environment and therefore happens at different $U_{\rm eff}$ values for the
three cations.   

\begin{figure}
\includegraphics{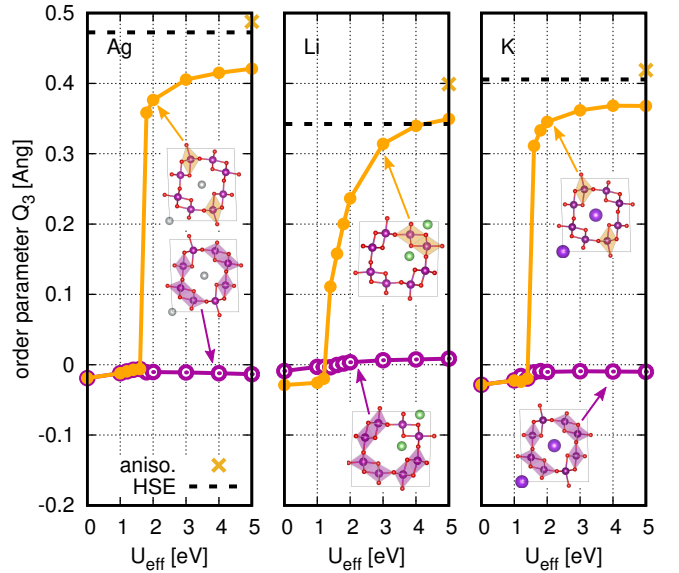}
\caption{\label{fig:CationIndex}
(color online) Averaged order parameter $Q_3$ as a function of $U_{\rm eff}$
with OptB88+$U_{\rm eff}$,
calculated for Mn$^{3+}$ centered octahedra (points) and
Mn$^{4+}$ centered octahedra (circles) for each of Ag, K and Li in A$_2$Mn$_8$O$_{16}$.
Results for HSE and PBE+($U$ = 6 eV, $J$ = 1 eV) are also shown.
Insets illustrate the charge order and provide a key to the octahedra sampled in each case.}
\end{figure}

\begin{figure}
\includegraphics{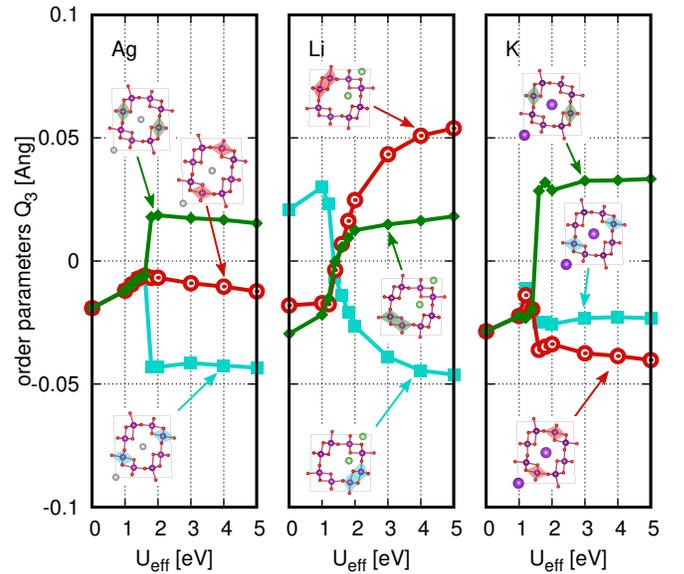}
\caption{\label{fig:CationIndexAll}
(color online) Same as in \figref{fig:CationIndex} but focusing on the order parameters $Q_3$ 
averaged in pairs for the six Mn$^{4+}$ octahedra, following the illustrative insets.}
\end{figure}

Quantitatively, we find that the local Jahn-Teller effect in the
A$_2$Mn$_8$O$_{16}$ hollandite materials can be specifically tracked by means of
the local order parameter $Q_3$, defined in \eqref{eq:Index}.  In those stable
structures in which local Mn$^{3+}$ ions emerge, correspondingly the order
parameter $Q_3$ becomes non-zero.  For the case at hand, full localization of
the two extra donated electrons per unit cell correspond to two Mn$^{3+}$
centered octahedra and six Mn$^{4+}$ centered octahedra.  In each case, the
octahedra naturally pair, with approximately similar distortions.  To highlight
the emergence of the Mn$^{3+}$ centers, we plot two averaged $Q_3$ parameters,
one for the two Mn$^{3+}$ ions and one for the remaining six Mn$^{4+}$
ions. The results for all three cation doped hollandite unit cells as a function
of $U_{\rm eff}$ are given in \figref{fig:CationIndex}. The insets illustrate the
charge order that emerges in each case.  The order parameter of the averaged six
Mn$^{4+}$ ions can be further decomposed into three pairs, as shown in
\figref{fig:CationIndexAll}.  While these three pairs of octahedra differ in
minor details of structure, the magnitude of $Q_3$ remains near zero.

Looking at \figref{fig:CationIndex}, one sees that, for all three compounds,
$Q_3$ increases significantly between $U_{\rm eff}=1$ and $2$ eV and indicates
the formation of Mn$^{3+}$. The charge
order has a different character in the case of Li, where as previously noted,
the small ionic radius favors an off-center ion location.  For large $U_{\rm
eff}$, this results in a local complex of two Li ions and two Mn$^{3+}$ centered
octahedra in a single tunnel wall. This is distinct from the local structure found
for $U_{\rm eff}=0$ and described above.  In contrast, for the K and Ag cases, where
the cation remains near the tunnel center, the two Mn$^{3+}$ centered octahedra
are maximally separated.  For reference, results based on HSE are also plotted.
The values track the large $U_{\rm eff}$ plateaux.
The anisotropic PBE+($U$=6, $J$=1 eV) predicts a larger distortion,
but also in overall good agreement witht the HSE results.
These results all correspond to AF order.
For comparison, in the Ag case, with FM order and $U_{\rm eff}=5$ eV,
$Q_3$ is a bit smaller, 0.38 \AA, indicating some interplay with the magnetic order.

\begin{figure}
\includegraphics{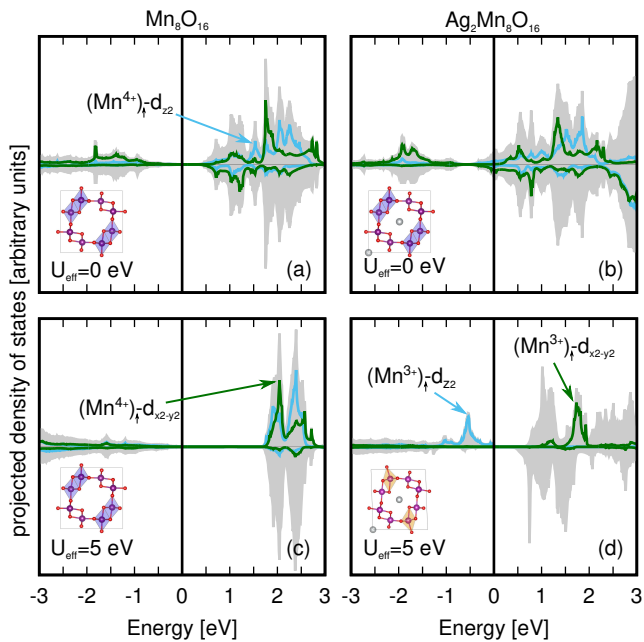}
\caption{\label{fig:egDOS}
Projected density of Mn $e_{g}$ states ($d_{z^2},d_{x^2-y^2}$) for
Mn$_8$O$_{16}$ (Fig. (a),(c)) and Ag$_2$Mn$_8$O$_{16}$ (Fig. (b),(d)). 
Grey background represents the $e_g$ states of all octahedra,
while the colored lines in all figures represent the labeled projections
of the majority spin up octahedra highlighted in the insets. 
}
\end{figure}

To understand this effect from the electronic structure perspective we have used
Wannier orbitals and calculated the projected density of states (PDOS) for all
Mn ions individually. To probe the local states in each octahedron in the most
natural way, we have aligned locally the $(x,y,z)$ axes as depicted in
\figref{fig:LocalWannier}.  Physically, it is sufficient to restrict
consideration to the local $e_g$ derived empty states that receive the electrons
from the A cations.  In \figref{fig:egDOS}, we compare the
$e_g$ derived PDOS for the case of pure Mn$_8$O$_{16}$ with that of
Ag$_2$Mn$_8$O$_{16}$ for PBE with $U_{\rm eff}=0$ and with $U_{\rm eff}=5$ eV.
For each case, the results for an AF ordered state (\figref{fig:UnitCell}) are shown
with the majority spin PDOS plotted upwards and the minority spin
PDOS plotted downwards.  The gray background shows the sum of all $e_g$ 
states.  The lines represent the $d_{z^2}$ (blue) and $d_{x^2-y^2}$ states
(green) from specific octahedra with majority spin up (see insets).

In the case of pristine hollandite, the results shown
are for a tetragonal cell and all eight octahedra are equivalent, except for the majority spin orientation.
The increase in
the interaction strength increases the band gap, as expected.  Correspondingly,
the hybridization that leads to a small amount of $d_{z^2}$ and $d_{x^2-y^2}$
character in the occupied portion of the PDOS, is reduced by the large $U_{\rm
eff}$. But effectively, all Mn ions in Mn$_8$O$_{16}$ have a $t_{2g}$ shell
locally filled with majority spin electrons, but empty for minority spin
electrons and no occupancy of the nominal $e_g$ bands. The Jahn-Teller effect is
absent.

For Ag$_2$Mn$_8$O$_{16}$, the additional $s$ electrons from Ag partially
occupy available $e_g$ majority spin states.  However, for $U_{\rm eff}=0$,
the system behaves like a band metal with partial occupation of several
previously empty bands and the eight octahedra remain equivalent. The PDOS in
\figref{fig:egDOS} (b) shows that the Fermi energy has moved into the bottom
portion of the previously empty $e_g$ derived bands and there is some detailed
rearrangement of states.  However,  \figref{fig:egDOS} (d) clearly shows that,
for a high value of $U_{\rm eff}=5$ eV, 
the majority $d_{z^2}$ state is fully occupied in two of
the identifiable local Mn$^{3+}$ octahedra.  
Correspondingly, the local $d_{x^2-y^2}$ state is empty
and a band gap is maintained.  
This large energy splitting goes with the large, local value of the $Q_3$ order
parameter illustrated in \figref{fig:CationIndex} and drives the energy gain
from the Jahn-Teller effect.  For the other six Mn$^{4+}$ octahedra, the $e_g$
states remain empty and the local value of the $Q_3$ order parameter is close to
zero.  Overall, the consequence is that the unit cell becomes monoclinic for
high $U_{\rm eff}$ due to the elongation of the Mn$^{3+}$O$_6$ octahedra in the
$b$ direction.  

\begin{figure}
\includegraphics{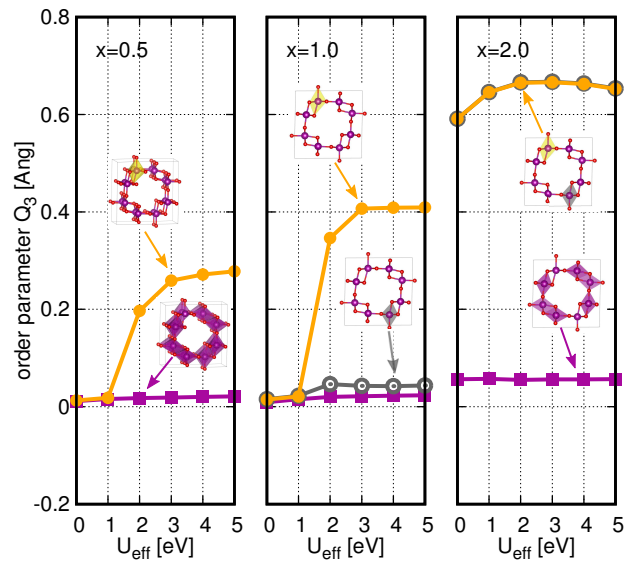}
\caption{\label{fig:JelliumQ3} Distortion index $Q_3$ as function of effective
parameter $U_{\rm eff}$ for charged unit cells (Mn$_8$O$_{16}$)$^{-x}$, as
defined in \eqref{eq:Index}.}
\end{figure}

In order to isolate the electronic effect from the ionic effect, we have also
done self consistent calculations in which electrons have been added to pristine
Mn$_8$O$_{16}$, compensated by a uniform positive background.  This allows us to
assess the electronic part of the Jahn-Teller effect in a continuous way as a
function of added electrons to the unit cell, in essence studying
(Mn$_8$O$_{16}$)$^{-x}$.  The results for three examples, using Opt-B88 and
different $U_{\rm eff}$ values, are shown in \figref{fig:JelliumQ3}. 

For a low charge of $x=0.5$ and the restriction of a single unit cell a charge
localized solution starts to emerge for $U_{\rm eff}\approx3$ eV, with a local
order parameter $Q_3$ that grows to a modest value ($Q_3=0.17$ \AA) for larger
$U_{\rm eff}$ (not shown). If a $1\times1\times2$ supercell is used the order
parameter $Q_3$ increases roughly by a factor of two
(see \figref{fig:JelliumQ3}). In addition, the magnetic moment of the
corresponding Mn ion increases from $m=3.4$ $\mu_B$ to $m=3.7$ $\mu_B$
indicating formation of a single Mn$^{3+}$ ion for the supercell.  For
$x=1.0$ a single Mn$^{3+}$ ion with a magnetic moment of $m=3.7 \mu_B$ starts
forming from a small value of $U_{\rm eff}$.  Finally, for the case of $x=2.0$,
which corresponds to the same number of electrons donated in
(Ag,K,Li)$_2$Mn$_8$O$_{16}$, the localized solutions already emerge with $U_{\rm
eff}=0$. Furthermore, the saturated value of the distortion for larger $U_{\rm
eff}$ exceeds that found for the real materials in \figref{fig:CationIndex}.
Together, these results suggest that the pure electronic driving force for
charge localization and Jahn-Teller distortion is quite strong.  However, along
with these local effects, the overall lattice is expanded by roughly 4\%.
Evidently, the presence of the actual cations in the tunnels (instead of a
simple neutralizing background) opposes this expansion and limits the
Jahn-Teller distortion.  

\subsection{Charge Ordering}\label{sec:ChargeOrdering}

Having demonstrated the emergence of solutions with localized Mn$^{3+}$ centered
octahedra due to the Jahn-Teller effect, we further examine the impact of how
those centers are ordered and the interplay with the magnetic order.  We focus
on Ag$_2$Mn$_8$O$_{16}$.  While a systematic study that extends beyond the basic
unit cell is desirable, the computationally complexity grows rapidly.  In
particular, the number of distinct (Mn$^{3+}$)$_2$(Mn$^{4+}$)$_6$
configurations within even modest sized super cells, assessed using site
occupancy disorder tool of Grau-Crespo,\cite{SOD2007} become quite
large.\footnote{ For example, we found 79 symmetrically distinct ways to
distribute two Mn$^{3+}$ in the $1\times1\times2$ super cell and more than 2900
configurations for a $1\times1\times3$ super cell.} Therefore, we have
approached this problem empirically, identifying the smallest building blocks of
Mn$^{3+}$ patterns and building more complex structures based on them. 

\begin{figure}
\includegraphics{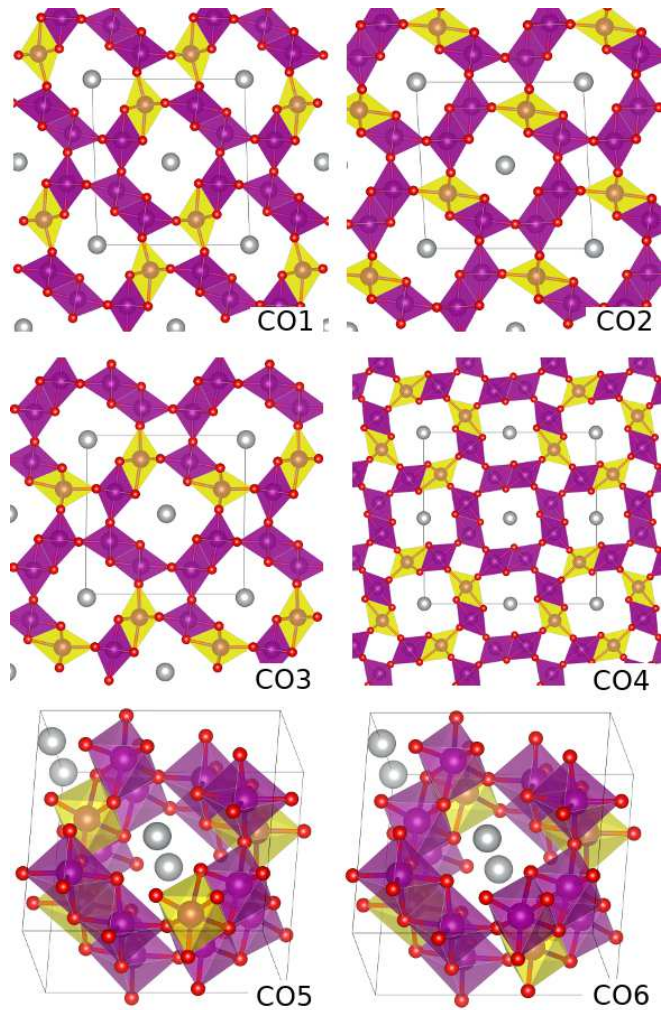}
\caption{\label{fig:Mn3Positions}
(color online) Distinct (Mn$^{3+}$)$_2$(Mn$^{4+}$)$_6$ charge
orderings  in A$_2$Mn$_8$O$_{16}$  investigated in this work. 
Yellow octahedra indicate Mn$^{3+}$ ions and
magenta octahedra indicate Mn$^{4+}$ ions.  Magnetic order is not shown.}
\end{figure}

To this end, we have assumed that only $d_{z^2}$ orbitals 
become occupied during charge localization, {\em i.e.}, 
elongation of Mn$^{4+}$O$_6$ octahedra occurs only in
the $(a,b)$ plane (see the Wannier projection in \figref{fig:LocalWannier}). This
gives three different charge orderings for the unit cell:
CO1, CO2, and CO3, visualized in \figref{fig:Mn3Positions}.\footnote{The
remaining two charge orderings for the unit cell are not stable under ionic
force minimization for both AF and FM spin alignment.} These orderings are
the building blocks for three more extended charge orderings
in supercells: CO4 ($\sqrt2\times\sqrt2\times1$) and CO5 and CO6 ($1\times1\times2$).
With the exception of CO3, this set coincides with the spin-charge ordering structures
studied by Fukuzawa {\em et al.} by means of unrestricted
Hartree-Fock calculations for K$_2$Mn$_8$O$_{16}$.\cite{Fukuzawa13} 
For our initial survey,
we have used PBE+$U_{\rm eff}$=5 eV, which combines rough agreement with the HSE
for binding energy of Ag$_2$Mn$_8$O$_{16}$ and clear formation of local Mn$^{3+}$ centers.
The results are
given in \tabref{tab:ChargeOrder}. 
 \begin{table}
 \caption{\label{tab:ChargeOrder}Energy (in eV per unit cell, relative to the
lowest energy case), magnetic ordering and lattice constants (in \AA) of
Ag$_2$Mn$^{3+}_2$Mn$_6^{4+}$O$_{16}$ configurations of \figref{fig:Mn3Positions}
obtained with several different exchange-correlation approximations.}
 \begin{ruledtabular}
 \begin{tabular}{l r r r r r r }
  conf.                & 
$E$   & 
    mag.   &
$a$ [\AA]  &
$b$ [\AA]  &
$c$ [\AA]  &
$\gamma$ [$^\circ$] \\
\hline 
\multicolumn{7}{c}{PBE+$U_{\rm eff}$=5}\\
 CO1 & 0.00 & FM &  9.55 & 10.32 & 2.93 & 91.14\\ 
 CO2 & 0.04 & FM &  9.91 &  9.90 & 2.94 & 93.58\\ 
 CO1 & 0.06 & AF &  9.54 & 10.35 & 2.92 & 91.49\\ 
 CO6 & 0.13 & FM &  9.85 &  9.85 & 5.92 & 90.02\\
 CO4 & 0.16 & FM & 13.93 & 13.97 & 2.96 & 90.21\\ 
 CO2 & 0.26 & AF &  9.80 & 10.06 & 2.92 & 93.55\\ 
 CO4 & 0.45 & AF & 14.06 & 13.99 & 2.92 & 90.25\\ 
 CO6 & 0.45 & AF &  9.88 &  9.88 & 5.85 & 90.01\\
 CO3 & 0.55 & AF &  9.91 &  9.94 & 2.92 & 89.55\\ 
 CO5 & 0.55 & AF &  9.88 &  9.89 & 5.85 & 90.00\\ 
\hline
\multicolumn{7}{c}{PBE+($U$=6, $J$=1 eV)}\\
 CO1 & 0.00 & AF &  9.53 & 10.35 & 2.90 & 91.38\\ 
 CO1 & 0.12 & FM &  9.53 & 10.36 & 2.91 & 91.30\\ 
 CO2 & 0.17 & AF &  9.78 & 10.06 & 2.91 & 93.38\\ 
 CO2 & 0.21 & FM &  9.79 & 10.05 & 2.92 & 93.30\\ 
 CO6 & 0.38 & AF &  9.87 &  9.88 & 5.83 & 90.02\\ 
 CO4 & 0.38 & AF & 13.98 & 14.06 & 2.91 & 90.25\\ 
 CO4 & 0.39 & FM & 13.98 & 14.05 & 2.92 & 90.25\\ 
 CO6 & 0.39 & FM &  9.87 &  9.88 & 5.86 & 90.03\\ 
 CO3 & 0.48 & AF &  9.90 &  9.94 & 2.91 & 90.46\\ 
 CO5 & 0.49 & FM &  9.87 &  9.87 & 5.86 & 90.00\\ 
 CO5 & 0.50 & AF &  9.87 &  9.88 & 5.83 & 90.00\\ 
\hline
\multicolumn{7}{c}{HSE}\\
 CO1 & 0.00 & AF & 9.41 & 10.10 & 2.84 & 91.64 \\
 CO2 & 0.24 & AF & 9.67 &  9.74 & 2.84 & 93.25 \\
 CO1 & 0.26 & FM & 9.39 & 10.10 & 2.84 & 91.63 \\
 CO2 & 0.38 & FM & 9.70 &  9.74 & 2.84 & 93.45 \\
 \hline
\multicolumn{7}{c}{Opt-B88}\\
Uniform     & 0.00 & AF & 9.66 & 9.66 & 2.84 & 90.00  \\
Uniform     & 0.67 & FM & 9.66 & 9.66 & 2.86 & 90.00  \\
\hline
\multicolumn{7}{c}{Opt-B88+$U_{\rm eff}$=1.6}\\
Uniform     & 0.00 & AF & 9.71& 9.72 & 2.84 & 90.19   \\
       CO1  & 0.02 & AF & 9.40&10.18 & 2.87 & 91.27   \\
Uniform     & 0.27 & FM & 9.73& 9.73 & 2.92 & 90.00   \\
 \end{tabular}
 \end{ruledtabular}
 \end{table}

Several competing phases with
different charge and magnetic order emerge.  The lowest pair differ by only 0.04 eV per
cell, but exhibit distinctive structures.  CO1 and CO2 both have ferromagnetic
(FM) order, but a different organization of the charge. CO1 has a strong
monoclinic distortion, whereas CO2 is almost perfectly orthorhombic with an
angle of $\gamma=93.58^\circ$ between $a$ and $b$. The next phase is the AF
ordered CO1 structure, 0.06 eV above the FM CO1 structure.  The charge order
CO6, 0.13 eV above FM CO1, is essentially tetragonal in structure, achieved by
organizing the Mn$^{3+}$ evenly around the tunnel walls, but alternating along
the $c$ axis in a $1\times1\times2$ unit cell.

The next distinct configuration, CO4 coincides formally with the charge
ordering found in K$_2$V$_8$O$_{16}$ at low temperatures.\cite{Isobe06} However,
the Jahn-Teller effect in Mn$^{3+}$ ions of CO4 acts in the $(a,b)$ plane (see
\figref{fig:LocalWannier} and \figref{fig:Mn3Positions}) in contrast to the
V$^{3+}$ ions, where the same effect is active primarily along the $c$ axis and
causes a dimerization of V$^{3+}$ (Peierls distortion) and a zig-zag pattern of
V$^{4+}$ along the tunnel direction.\cite{Kim16} On the other hand, the
distortion modes in Ag$_2$Mn$_8$O$_{16}$ compensate each other and yield
effectively a tetragonal structure for both, the ferromagnetic and
antiferromagnetic ordering. 
Several other choices of charge and magnetic order also yield stable solutions,
but with higher energy (\tabref{tab:ChargeOrder}).  

With PBE+($U$=6, $J$=1 eV), a similar picture emerges (\tabref{tab:ChargeOrder}).
The lowest energy structure has
a strong monoclinic distortion, but several structures close in energy exhibit
smaller distortions.  
However, the interplay with the magnetic order is different.
The lowest energy phase with charge order CO1 and CO2 both have AF order.
Limited exploration with the hybrid HSE clearly confirms the charge ordered
ground state. The results for CO1 show a similar degree of monoclinic
distortion. The energy separation to the higher lying CO2 state is increased, 
but the nearly tetragonal structure is similar to that found 
with PBE+$U_{\rm eff}$=5 eV and shows a similar value of the $Q_3$ order parameter (see
\figref{fig:CationIndex}). 
Furthermore, the magnetic order found with HSE agrees with
that from PBE+($U$=6, $J$=1 eV) and the energy difference
between the higher FM phase and the AF phase is about the same.

For comparison, the Opt-B88 results
show a uniform phase, as expected for
a band metal with no charge localization, a tetragonal structure
and a much larger stabilization of the AF order over the FM order.
Interestingly, inclusion of a small $U_{\rm eff}$ does allow for a charge ordered structure
which is only slightly higher in energy than the uniform structure
and reduces the energy difference separating the FM ordered phase.

For reference, the predicted magnetic order for the parent $\alpha-$MnO$_2$ using
Opt-B88 and PBE is AF, with an energy difference per unit cell of 0.70 and
0.58 eV to the FM phase.  Explicit consideration of local Coulomb interactions
through GGA+$U_{\rm eff}$ systematically reduces this energy splitting, to 0.01 and 0.06
eV respectively for $U_{\rm eff}$=5 eV. The same trend was found in
Ref.~\onlinecite{Crespo13a}, but we do not find stabilization of the FM phase in
the range of  $U_{\rm eff}$ we considered. 
With PBE+($U$=6, $J$=1 eV), the AF-FM energy splitting is increased (0.14 eV).  
Finally, our results for HSE also indicate AF order to be favored, with a
similar splitting (0.24 eV). While internally consistent, this final result
for $\alpha-$MnO$_2$ is opposite to that found in Ref.~\onlinecite{Crespo13a}. We
have included the semicore electrons and the structure for each phase was
relaxed with HSE.

Finally, we note that our HSE calculations
also yield an AF phase for the K and Li cases
with an energy splitting of 0.20 and 0.14 eV respectively to the FM phase.
Interestingly, PBE+($U$=6, $J$=1 eV) results show an AF phase for K (0.12 eV),
but a FM phase for Li (0.06 eV).

\section{Discussion}\label{sec:Discussion}

Our results based on PBE+$U_{\rm eff}$ at high $U_{\rm eff}$ values show a charge
ordered ground state, but several close-by competing charge and magnetic ordered
phases.  The HSE results confirm the charge ordering in the ground state.
However, also considering the PBE+($U$=6, $J$=1 eV) results,
the details of the interplay between charge order and magnetic coupling
are sensitive to the method chosen. More generally, these results,
particularly the low energy scale, point to the idea that at higher temperature,
disordered phases may be more relevant to the measured room temperature results.
For context, we briefly review what is known about exemplary, related compounds.

Near stoichiometric K$_2$V$_8$O$_{16}$ was obtained through high pressure
synthesis.\cite{Isobe06} While exhibiting the ideal, tetragonal hollandite
structure at room temperature, a first order transition to a monoclinic phase
was observed at 170 K, with a corresponding change from conducting to insulating
behavior.  Subsequent X-ray and neutron diffraction studies clearly showed that
the monoclinic low temperature phase consisted of charge-ordered, localized
V$^{3+}$ in a $\sqrt{2}\times\sqrt{2}\times2$ supercell.\cite{Komarek11}
Detailed calculations based on GGA+$U_{\rm eff}$=3 eV provided a corresponding picture of the
electronic structure. While local Jahn-Teller effects manifested for both
V$^{3+}$ and V$^{4+}$ centered octahedra, strong, local antiferromagnetic
coupling between dimers along the chains in the tunnel direction drove the
opening of the energy gap and the metal to insulator transition.\cite{Kim16} 
The reduced, monoclinic symmetry at low temperature was secondary. This agreed
with the observation that Rb$_2$V$_8$O$_{16}$ shows a metal-insulator
transition, but remains tetragonal.\cite{Isobe09}

The strong coupling of V$^{3+}$ along the $c$ direction can be related to the
partially filled $t_{2g}$ shell which is Jahn-Teller active. In contrast, it can
be expected that the crystal distortions in hollandite chromates are weak due to
the full subshell in Cr$^{3+}$. This was also shown for K$_2$Cr$_8$O$_{16}$,
where ferromagnetic order emerged at 180 K, followed by a metal to insulator
transition at 95 K (with ferromagnetic order maintained).\cite{Hasegawa09} While
no structural transition was initially observed, a subsequent study detected a
subtle, $\sqrt{2} \times \sqrt{2}$ monoclinic phase at low temperature
($P112_1/a$) with no measurable distortion of the lattice parameters, but small
dimerization in the Cr-O bond lengths (scale 0.03 \AA).\cite{Toriyama11}
However, bond valence sum analysis suggested no charge localization, that is, no
evidence for local Cr$^{3+}$ ions.  
GGA+$U_{\rm eff}$=2.9 eV calculations showed that the
small dimerization was sufficient to open a small gap, while remaining
consistent with the ferromagnetic order. 
A separate GGA+$U_{\rm eff}$=3 eV based study
identified a specific soft mode associated with this Peierls driven
transition.\cite{Kim14} 

The situation for hollandite group manganese oxides under study here has been
much less clear.  Early studies of the parent $\alpha-$MnO$_2$ showed
antiferromagnetic order below 24.5 K.\cite{Yamamoto74} Studies of
K$_x$Mn$_8$O$_{16}$ with $x$ near 1.5 showed a change in the activation energy
associated with electrical conductivity around 200 K and changes in magnetic
susceptibility at low temperature with some ambiguity as to the details of the
ordering.\cite{Strobel84,Sato99} A study examining K concentration dependence of
the magnetic ordering ($0.8 \leq x \leq 1.4$) suggested a spin glass for lower
concentration and antiferromagnetic order for higher concentration.\cite{Luo10}
Recently, a study for samples with $x$ about 1.2 and 1.4 found that even for
these concentrations, a spin glass was more likely at low
temperature.\cite{Barudzija16} At room temperature, structural characterization
of these samples points to a tetragonal structure ($I4/m$).  Turning to other
examples, the original report of Ag$_{1.8}$Mn$_8$O$_{16}$ synthesis and
characterization reported a tetragonal structure.\cite{Jansen1984} On the other
hand, Ba$_{1.2}$Mn$_8$O$_{16}$ was found to have a monoclinic structure ($I2/m$)
already at room temperature and weak ferromagnetic order (or more complex competing
interactions) below 40 K.\cite{Ishiwata06} While some researchers have proposed
localization of Mn$^{3+}$ ions,\cite{Fukuzawa13} there has been no compelling
evidence. In particular, detailed structural studies at low temperatures have
not been reported.

The chemically related system LiMn$_2$O$_4$, which would also be expected to
show a mixed valence state between Mn$^{4+}$ and Mn$^{3+}$, exhibited a first
order transition near 280 K to a structure with clear evidence of charge and
orbital ordering driven by the Jahn-Teller effect.\cite{RodriguezCarvajal98,
Massarotti99} Calculations based on GGA+$U_{\rm eff}$ = 4.5 eV showed localized Mn$^{3+}$
ions with corresponding Jahn-Teller distortion.\cite{Ouyang09}  The computed
lattice parameters were also in good agreement with those measured for the low
temperature, ordered phase.

The role of ionic radius has been analyzed early-on based on an extensive set of
hollandite group samples, with the criterion for monoclinic distortion deduced
to be $r_{\rm A} < r_{\rm Mn}/0.48$.  Using Shannon ionic radii (Mn$^{4+}$ = 0.53 \AA~
(6-fold), Li$^{1+}$ = 0.59 \AA~(4-fold) and Ba$^{2+}$ = 1.42
\AA),\cite{Shannon76} this criterion predicts that Ba$_2$Mn$_8$O$_{16}$ should
be tetragonal and that Li$_2$Mn$_8$O$_{16}$ should be monoclinic.  The data for
the ion exchanged samples that show a tetragonal structure for the Li case
involve extra factors due to the role of water,\cite{Kadoma2007} and may not be
a definitive test. In the Ba case, room temperature measurements already show a
mildly monoclinic structure,\cite{Ishiwata06} suggesting that other factors are
likely involved.

The detailed understanding of the V and Cr based hollandites and
the mixed valence LiMn$_2$O$_4$ was fully supported by GGA+$U_{\rm eff}$
calculations which explicitly took into account local Coulomb interactions
among the transition metal $3d$ electrons.
The range of $U_{\rm eff}$ values used were similar to those in which
our study showed formation of local Mn$^{3+}$ centers
in the A$_2$Mn$_8$O$_{16}$ systems and relatively accurate
binding energy for the A atoms in the tunnels.
Our reference HSE calculations also clearly showed charge localization.
Finally, the chemically closest analog for which clear temperature dependent structural
measurements were available,  LiMn$_2$O$_4$, showed a charge ordered phase at low temperature.
Taken together, this evidence indicates that charge localization in the
A$_2$Mn$_8$O$_{16}$ system, to form Mn$^{3+}$ centers is likely the correct physical
picture.    

The proper interpretation of the room
temperature structural data remains an open question.  One possibility would be
that we simply have not yet identified the supercell with a robust charge order
and a (near) tetragonal unit cell. We can not rule this out.  The other
possibility is that room temperature measurements are probing a disordered
phase, relative to the localized Mn$^{3+}$ centers.  This picture is supported
by our PBE+$U_{\rm eff}=5$ eV and PBE+($U$=6, $J$=1 eV) results reported in
\tabref{tab:ChargeOrder} for the Ag$_2$Mn$_8$O$_{16}$ case. A series of
charge ordered phases are separated by small energy differences and encompass
phases with near tetragonal structure.  While the Li case appears to be more
complicated, with the fundamental unit emerging as two Li ions and two
edge-sharing Mn$^{3+}$ octahedra, the same basic picture could hold.

The magnetic order and the details of the energy differences between
the competing phases has emerged as a more subtle problem.  As already discussed in
Ref. \onlinecite{Crespo13a}, the competing effects that control the magnetic
coupling between neighboring Mn ions are quite subtle for the hollandite
structure, particularly the coupling across the corner shared sites.  The
broadly applicable Goodenough-Kanamori-Anderson (GKA) rules clearly distinguish
over two scenarios: Mn-O-Mn links with near 90$^\circ$ and near 180$^\circ$ bond
angles on the connecting O center, favoring FM and AF coupling
respectively.\cite{Anderson50, Goodenough55,Kanamori1960} In hollandite, the
Mn-O-Mn bond between two tunnel walls is roughly 130$^\circ$, indicating that
the balance of interactions should be expected to be quite delicate.  The
differences we find likely reflect tipping this balance one way or the other.
The complexity of the observed magnetic signatures in experiment, described
above, similarly indicate relatively weak magnetic coupling, possibly sensitive
to other factors in the material composition or structure.  The hollandites
represent challenging systems for DFT-based methods.
While our results based on HSE and PBE+$(U=6,J=1)$ eV are rather consistent,
the energy scale of the differences can be significant compared to
the temperature scale, discussed above, at which phase transitions are observed in related compounds. 

This study was motivated by the need to calibrate DFT-based methods, inherently
applicable at T=0, for use in studies of the broader phase diagram associated
with Li$_x$Ag$_y$Mn$_8$O$_{16-z}$:(H$_2$O)$_w$.  Much of the associated
practical, electrochemistry research is based on room temperature
characterization.  The emergence of strong monoclinic distortion, already for
Ag$_2$Mn$_8$O$_{16}$, signaled a disconnect that required further investigation.
We have argued here that the formation of localized Mn$^{3+}$ centers upon
reduction by addition of Ag, Li or K, with the associated Jahn-Teller
distortions, is likely correct.  Parenthetically, we have found that the
potential surfaces that describe the resulting cell distortions can be rather
flat, making full convergence to the lowest energy structure challenging.  In
this picture, disorder at finite temperature explains the observation of higher
symmetry structures.  As a practical matter, calculations that fully account for
the complexity of this picture can not be used for phase diagram exploration.

Our results suggest two alternatives. Referring back to
\figref{fig:BindingEnergy}, the GGA+$U$ methods with moderate $U$ in the range of
3-5 eV give relatively accurate addition energies for Ag, Li and K.  
Furthermore, inclusion of the full anisotropy in the Coulomb interaction
improved the results. However,
these calculations, at T=0, will have charge ordered solutions with
corresponding distortions of structure.  This discrepancy from room temperature
structure must be acknowledged, but may not be of practical significance for
electrochemical properties, driven largely by energy.  Alternatively, Opt-B88
with small $U$, also gives relatively accurate addition energies.  Below the
thresholds identified in \figref{fig:CationIndex}, electronically driven charge
localization does not occur (as distinct from Li ion driven distortions).  For
mapping phase diagrams in this system, this alternative picture may also be
adequate for electrochemical properties.  It is also likely that a functional
with explicit inclusion of van der Waals interactions will better describe
H$_2$O in the tunnels.  However, there is a strong caveat: in this
approximation, the reduced system is modeled as a band metal and
correspondingly, properties such as optical conductivity may be fundamentally
wrong.  Finally, we note that some of the features of this phase diagram that we
have already described, Ag segregation upon addition of Li and the formation of
sheet-like phases with Li, emerge with very similar energetics using either of
these two approaches.\cite{Xu17}

\section{Conclusion}\label{sec:Conclusion} 

We have carefully mapped out the ground state phases of A$_2$Mn$_8$O$_{16}$ in
the hollandite structure for Ag, Li and K in the A position, as they are
predicted by DFT-based methods with several different approximate
exchange-correlation functionals.  These methods included PBE, Opt-B88
(explicitly including van der Waals interactions), their generalization to
approximately include local Coulomb interactions among the Mn $3d$ electrons (GGA+$U$
approach), the meta-GGA SCAN and finally an exemplary hybrid functional, the
HSE approach.

Using HSE as a reference for the A addition energy, we found that Opt-B88 with a
small value of $U$ was rather accurate across Ag, Li and K insertion.  SCAN was
also relatively accurate overall, but unfortunately predicted that Ag in
Ag$_2$Mn$_8$O$_{16}$ was not bound.  Furthermore, these methods predicted that
the reduced compounds were band metals, with a tetragonal unit cell for the Ag
and K cases.  This agrees with available, room temperature measurements of structure, but not
with the HSE calculations.  For Li, the small ionic radius generically led to
different local structures such that the ionic interactions with the MnO$_2$
walls drove local distortions and a monoclinic unit cell.  Interestingly, this
agrees with predictions of an empirical relationship derived from hollandite
mineral data, based on the ratio of the Li ionic radius to that of Mn.

Alternatively, PBE+$U$ with $U_{\rm eff}$ in the 3-5 eV range, showed similar
accuracy for the A addition energy. For this range of $U_{\rm eff}$, localization
of the electrons donated by Ag, Li and K to form specific Mn$^{3+}$ centers was
clearly observed. This agrees with the HSE calculations. The $Q_3$ order
parameter of Van Vleck was found to describe the associated local Jahn-Teller
distortion driven by the introduction of the extra $3d$ electron on Mn$^{3+}$ into
the previously empty $e_g$ derived states. The net result is strong monoclinic
distortion of the unit cell. This does not agree with available room
temperature structural data. However, further analysis of competing charge
ordered phases shows a series of phases with relatively small energy difference
from the predicted ground state phase. These include phases with nearly
tetragonal unit cells. Based on comparison with studies of the temperature
dependence of the properties of single crystal V and Cr based hollandites, as
well as the mixed valence LiMn$_2$O$_4$ spinel compound, we conclude that the
charge localization to form Mn$^{3+}$ centers is likely physically correct.
Then the observed high symmetry structures at room temperature
would reflect disorder among the Mn$^{3+}$ centers.

At the same energy scale, the interplay with magnetic order is significant.
The HSE calculations
predict an AF phase for the parent $\alpha-$MnO$_2$ 
and all three A$_2$Mn$_8$O$_{16}$ compounds studied.
Here, the best agreement was obtained 
employing the anisotropic PBE+($U$=6, $J$=1 eV) eV approach,
although curiously, not for the Li case.
This highlights the challenge in capturing all of the competing
effects in these materials with sufficient accuracy for detailed simulation
of the competing phases and transitions among them.

We have discussed the consequences of these results for use of these DFT-based
methods to survey the phase diagram of Li$_x$Ag$_y$Mn$_8$O$_{16-z}$ where a
large number of compounds and structures must be explored efficiently.  We have
outlined the trade-offs involved. Energetics may be captured with either low $U$
or high $U$ approaches.  However, the convenience of avoiding the Jahn-Teller
driven distortions will naturally mean sacrificing the ability to predict other
properties that are sensitive to charge localization, such as optical
conductivity.

\begin{acknowledgments}
The work was supported as part of the Center for Mesoscale Transport Properties
(m2M), an Energy Frontier Research Center funded by the U.S. Department of
Energy, Office of Science, Basic Energy Sciences, under Award \# DE-SC0012673.
Research done in part using facilities in the Center for Functional
Nanomaterials, which is a U.S. DOE Office of Science User Facility, at
Brookhaven National Laboratory under Contract No. DE-SC0012704,
the high-performance LI-red and Handy computing systems 
at the Institute of Advanced Computational Sciences (IACS) 
and the SeaWulf cluster at Stony Brook University. 
\end{acknowledgments}

\bibliography{references}

\begin{thebibliography}{93}%
\makeatletter
\providecommand \@ifxundefined [1]{%
 \@ifx{#1\undefined}
}%
\providecommand \@ifnum [1]{%
 \ifnum #1\expandafter \@firstoftwo
 \else \expandafter \@secondoftwo
 \fi
}%
\providecommand \@ifx [1]{%
 \ifx #1\expandafter \@firstoftwo
 \else \expandafter \@secondoftwo
 \fi
}%
\providecommand \natexlab [1]{#1}%
\providecommand \enquote  [1]{``#1''}%
\providecommand \bibnamefont  [1]{#1}%
\providecommand \bibfnamefont [1]{#1}%
\providecommand \citenamefont [1]{#1}%
\providecommand \href@noop [0]{\@secondoftwo}%
\providecommand \href [0]{\begingroup \@sanitize@url \@href}%
\providecommand \@href[1]{\@@startlink{#1}\@@href}%
\providecommand \@@href[1]{\endgroup#1\@@endlink}%
\providecommand \@sanitize@url [0]{\catcode `\\12\catcode `\$12\catcode
  `\&12\catcode `\#12\catcode `\^12\catcode `\_12\catcode `\%12\relax}%
\providecommand \@@startlink[1]{}%
\providecommand \@@endlink[0]{}%
\providecommand \url  [0]{\begingroup\@sanitize@url \@url }%
\providecommand \@url [1]{\endgroup\@href {#1}{\urlprefix }}%
\providecommand \urlprefix  [0]{URL }%
\providecommand \Eprint [0]{\href }%
\providecommand \doibase [0]{http://dx.doi.org/}%
\providecommand \selectlanguage [0]{\@gobble}%
\providecommand \bibinfo  [0]{\@secondoftwo}%
\providecommand \bibfield  [0]{\@secondoftwo}%
\providecommand \translation [1]{[#1]}%
\providecommand \BibitemOpen [0]{}%
\providecommand \bibitemStop [0]{}%
\providecommand \bibitemNoStop [0]{.\EOS\space}%
\providecommand \EOS [0]{\spacefactor3000\relax}%
\providecommand \BibitemShut  [1]{\csname bibitem#1\endcsname}%
\let\auto@bib@innerbib\@empty
\bibitem [{\citenamefont {Post}(1999)}]{Post99}%
  \BibitemOpen
  \bibfield  {author} {\bibinfo {author} {\bibfnamefont {J.~E.}\ \bibnamefont
  {Post}},\ }\bibfield  {title} {\enquote {\bibinfo {title} {Manganese oxide
  minerals: Crystal structures and economic and environmental significance},}\
  }\href@noop {} {\bibfield  {journal} {\bibinfo  {journal} {PNAS}\ }\textbf
  {\bibinfo {volume} {96}},\ \bibinfo {pages} {3447--3454} (\bibinfo {year}
  {1999})}\BibitemShut {NoStop}%
\bibitem [{\citenamefont {Bystrom}\ and\ \citenamefont
  {Bystrom}(1950)}]{Bystrom50}%
  \BibitemOpen
  \bibfield  {author} {\bibinfo {author} {\bibfnamefont {A.}~\bibnamefont
  {Bystrom}}\ and\ \bibinfo {author} {\bibfnamefont {A.~M.}\ \bibnamefont
  {Bystrom}},\ }\bibfield  {title} {\enquote {\bibinfo {title} {The crystal
  structure of hollandite, the related manganese oxide minerals, and
  $\alpha$-{Mn}{O}$_{2}$},}\ }\href@noop {} {\bibfield  {journal} {\bibinfo
  {journal} {Acta Crystallogr.}\ }\textbf {\bibinfo {volume} {3}},\ \bibinfo
  {pages} {146--154} (\bibinfo {year} {1950})}\BibitemShut {NoStop}%
\bibitem [{\citenamefont {Biagioni}\ \emph {et~al.}(2013)\citenamefont
  {Biagioni}, \citenamefont {Capalbo},\ and\ \citenamefont
  {Pasero}}]{Biagioni13}%
  \BibitemOpen
  \bibfield  {author} {\bibinfo {author} {\bibfnamefont {C.}~\bibnamefont
  {Biagioni}}, \bibinfo {author} {\bibfnamefont {C.}~\bibnamefont {Capalbo}}, \
  and\ \bibinfo {author} {\bibfnamefont {M.}~\bibnamefont {Pasero}},\
  }\bibfield  {title} {\enquote {\bibinfo {title} {Nomenclature tunings in the
  hollandite supergroup},}\ }\href@noop {} {\bibfield  {journal} {\bibinfo
  {journal} {Eur. J. Miner.}\ }\textbf {\bibinfo {volume} {25}},\ \bibinfo
  {pages} {85--90} (\bibinfo {year} {2013})}\BibitemShut {NoStop}%
\bibitem [{\citenamefont {Post}\ \emph {et~al.}(1982)\citenamefont {Post},
  \citenamefont {Dreele},\ and\ \citenamefont {Buseck}}]{Post82}%
  \BibitemOpen
  \bibfield  {author} {\bibinfo {author} {\bibfnamefont {J.~E.}\ \bibnamefont
  {Post}}, \bibinfo {author} {\bibfnamefont {R.~B.~Von}\ \bibnamefont
  {Dreele}}, \ and\ \bibinfo {author} {\bibfnamefont {P.~R.}\ \bibnamefont
  {Buseck}},\ }\bibfield  {title} {\enquote {\bibinfo {title} {Symmetry and
  cation displacements in hollandites: structure refinements of hollandite,
  cryptomelane and priderite},}\ }\href@noop {} {\bibfield  {journal} {\bibinfo
   {journal} {Acta Crystallogr. Sect. B}\ }\textbf {\bibinfo {volume} {38}},\
  \bibinfo {pages} {1056--1065} (\bibinfo {year} {1982})}\BibitemShut {NoStop}%
\bibitem [{\citenamefont {Brock}\ \emph {et~al.}(1998)\citenamefont {Brock},
  \citenamefont {Duan}, \citenamefont {Tian}, \citenamefont {Giraldo},
  \citenamefont {Zhou},\ and\ \citenamefont {Suib}}]{Brock98}%
  \BibitemOpen
  \bibfield  {author} {\bibinfo {author} {\bibfnamefont {S.~L.}\ \bibnamefont
  {Brock}}, \bibinfo {author} {\bibfnamefont {N.}~\bibnamefont {Duan}},
  \bibinfo {author} {\bibfnamefont {Z.~R.}\ \bibnamefont {Tian}}, \bibinfo
  {author} {\bibfnamefont {O.}~\bibnamefont {Giraldo}}, \bibinfo {author}
  {\bibfnamefont {H.}~\bibnamefont {Zhou}}, \ and\ \bibinfo {author}
  {\bibfnamefont {S.~L.}\ \bibnamefont {Suib}},\ }\bibfield  {title} {\enquote
  {\bibinfo {title} {A review of porous manganese oxide materials},}\
  }\href@noop {} {\bibfield  {journal} {\bibinfo  {journal} {Chem. Mater.}\
  }\textbf {\bibinfo {volume} {10}},\ \bibinfo {pages} {2619--2628} (\bibinfo
  {year} {1998})}\BibitemShut {NoStop}%
\bibitem [{\citenamefont {Beyeler}(1976)}]{Beyeler76}%
  \BibitemOpen
  \bibfield  {author} {\bibinfo {author} {\bibfnamefont {H.~U.}\ \bibnamefont
  {Beyeler}},\ }\bibfield  {title} {\enquote {\bibinfo {title} {Cationic
  short-range order in the hollandite
  {K}$_{1.54}${Mg}$_{0.77}${Ti}$_{7.23}${O}$_{16}$: Evidence for the importance
  of ion-ion interactions in superionic conductors},}\ }\href@noop {}
  {\bibfield  {journal} {\bibinfo  {journal} {Phys. Rev. Lett.}\ }\textbf
  {\bibinfo {volume} {37}},\ \bibinfo {pages} {1557--1560} (\bibinfo {year}
  {1976})}\BibitemShut {NoStop}%
\bibitem [{\citenamefont {Bernasconi}\ \emph {et~al.}(1979)\citenamefont
  {Bernasconi}, \citenamefont {Beyeler},\ and\ \citenamefont
  {Strassler}}]{Bernasconi79}%
  \BibitemOpen
  \bibfield  {author} {\bibinfo {author} {\bibfnamefont {J.}~\bibnamefont
  {Bernasconi}}, \bibinfo {author} {\bibfnamefont {H.~U.}\ \bibnamefont
  {Beyeler}}, \ and\ \bibinfo {author} {\bibfnamefont {S.}~\bibnamefont
  {Strassler}},\ }\bibfield  {title} {\enquote {\bibinfo {title} {Anomalous
  frequency-dependent conductivity in disordered one-dimesnsional systems},}\
  }\href@noop {} {\bibfield  {journal} {\bibinfo  {journal} {Phys. Rev. Lett.}\
  }\textbf {\bibinfo {volume} {42}},\ \bibinfo {pages} {819--822} (\bibinfo
  {year} {1979})}\BibitemShut {NoStop}%
\bibitem [{\citenamefont {Chen}\ \emph {et~al.}(2007)\citenamefont {Chen},
  \citenamefont {Tang}, \citenamefont {Liu}, \citenamefont {Zhan},
  \citenamefont {Li}, \citenamefont {Huang},\ and\ \citenamefont
  {Shen}}]{Chen07}%
  \BibitemOpen
  \bibfield  {author} {\bibinfo {author} {\bibfnamefont {J.~L.}\ \bibnamefont
  {Chen}}, \bibinfo {author} {\bibfnamefont {X.~F.}\ \bibnamefont {Tang}},
  \bibinfo {author} {\bibfnamefont {J.~L.}\ \bibnamefont {Liu}}, \bibinfo
  {author} {\bibfnamefont {E.~S.}\ \bibnamefont {Zhan}}, \bibinfo {author}
  {\bibfnamefont {J.}~\bibnamefont {Li}}, \bibinfo {author} {\bibfnamefont
  {X.~M.}\ \bibnamefont {Huang}}, \ and\ \bibinfo {author} {\bibfnamefont
  {W.~J.}\ \bibnamefont {Shen}},\ }\bibfield  {title} {\enquote {\bibinfo
  {title} {Synthesis and characterization of {Ag}-hollandite nanofibers and its
  catalytic application in ethanol oxidation},}\ }\href@noop {} {\bibfield
  {journal} {\bibinfo  {journal} {Chem. Mater.}\ }\textbf {\bibinfo {volume}
  {19}},\ \bibinfo {pages} {4292--4299} (\bibinfo {year} {2007})}\BibitemShut
  {NoStop}%
\bibitem [{\citenamefont {Dharmarathna}\ \emph {et~al.}(2012)\citenamefont
  {Dharmarathna}, \citenamefont {King'ondu}, \citenamefont {Pedrick},
  \citenamefont {Pahalagedara},\ and\ \citenamefont {Suib}}]{Dharmarathna12}%
  \BibitemOpen
  \bibfield  {author} {\bibinfo {author} {\bibfnamefont {S.}~\bibnamefont
  {Dharmarathna}}, \bibinfo {author} {\bibfnamefont {C.~K.}\ \bibnamefont
  {King'ondu}}, \bibinfo {author} {\bibfnamefont {W.}~\bibnamefont {Pedrick}},
  \bibinfo {author} {\bibfnamefont {L.}~\bibnamefont {Pahalagedara}}, \ and\
  \bibinfo {author} {\bibfnamefont {S.~L.}\ \bibnamefont {Suib}},\ }\bibfield
  {title} {\enquote {\bibinfo {title} {Direct sonochemical synthesis of
  manganese octahedral molecular sieve ({OMS}-2) nanomaterials using cosolvent
  systems, their characterization, and catalytic applications},}\ }\href@noop
  {} {\bibfield  {journal} {\bibinfo  {journal} {Chem. Mater.}\ }\textbf
  {\bibinfo {volume} {24}},\ \bibinfo {pages} {705--712} (\bibinfo {year}
  {2012})}\BibitemShut {NoStop}%
\bibitem [{\citenamefont {Rossouw}\ \emph {et~al.}(1992)\citenamefont
  {Rossouw}, \citenamefont {Liles}, \citenamefont {Thackeray}, \citenamefont
  {David},\ and\ \citenamefont {Hull}}]{ROSSOUW1992}%
  \BibitemOpen
  \bibfield  {author} {\bibinfo {author} {\bibfnamefont {M.~H.}\ \bibnamefont
  {Rossouw}}, \bibinfo {author} {\bibfnamefont {D.~C.}\ \bibnamefont {Liles}},
  \bibinfo {author} {\bibfnamefont {M.~M.}\ \bibnamefont {Thackeray}}, \bibinfo
  {author} {\bibfnamefont {W.~I.~F.}\ \bibnamefont {David}}, \ and\ \bibinfo
  {author} {\bibfnamefont {S.}~\bibnamefont {Hull}},\ }\bibfield  {title}
  {\enquote {\bibinfo {title} {Alpha manganese dioxide for lithium batteries: A
  structural and electrochemical study},}\ }\href@noop {} {\bibfield  {journal}
  {\bibinfo  {journal} {Mater. Res. Bull.}\ }\textbf {\bibinfo {volume} {27}},\
  \bibinfo {pages} {221 -- 230} (\bibinfo {year} {1992})}\BibitemShut {NoStop}%
\bibitem [{\citenamefont {Johnson}\ \emph {et~al.}(1997)\citenamefont
  {Johnson}, \citenamefont {Mansuetto}, \citenamefont {Thackeray},
  \citenamefont {Shao-Horn},\ and\ \citenamefont {Hackney}}]{Johnson97a}%
  \BibitemOpen
  \bibfield  {author} {\bibinfo {author} {\bibfnamefont {C.~S.}\ \bibnamefont
  {Johnson}}, \bibinfo {author} {\bibfnamefont {M.~F.}\ \bibnamefont
  {Mansuetto}}, \bibinfo {author} {\bibfnamefont {M.~M.}\ \bibnamefont
  {Thackeray}}, \bibinfo {author} {\bibfnamefont {Y.}~\bibnamefont
  {Shao-Horn}}, \ and\ \bibinfo {author} {\bibfnamefont {S.~A.}\ \bibnamefont
  {Hackney}},\ }\bibfield  {title} {\enquote {\bibinfo {title} {Stabilized
  alpha-{Mn}{O}$_{2}$ electrodes for rechargeable 3 {V} lithium batteries},}\
  }\href@noop {} {\bibfield  {journal} {\bibinfo  {journal} {J. Electrochem.
  Soc.}\ }\textbf {\bibinfo {volume} {144}},\ \bibinfo {pages} {2279--2283}
  (\bibinfo {year} {1997})}\BibitemShut {NoStop}%
\bibitem [{\citenamefont {Dai}\ \emph {et~al.}(2000)\citenamefont {Dai},
  \citenamefont {Li}, \citenamefont {Siow},\ and\ \citenamefont {Gao}}]{Dai00}%
  \BibitemOpen
  \bibfield  {author} {\bibinfo {author} {\bibfnamefont {J.}~\bibnamefont
  {Dai}}, \bibinfo {author} {\bibfnamefont {S.~F.~Y.}\ \bibnamefont {Li}},
  \bibinfo {author} {\bibfnamefont {Kok~S.}\ \bibnamefont {Siow}}, \ and\
  \bibinfo {author} {\bibfnamefont {Z.}~\bibnamefont {Gao}},\ }\bibfield
  {title} {\enquote {\bibinfo {title} {Synthesis and characterization of the
  hollandite-type {Mn}{O}$_{2}$ as a cathode material in lithium batteries},}\
  }\href@noop {} {\bibfield  {journal} {\bibinfo  {journal} {Electrochim.
  Acta}\ }\textbf {\bibinfo {volume} {45}},\ \bibinfo {pages} {2211--2217}
  (\bibinfo {year} {2000})}\BibitemShut {NoStop}%
\bibitem [{\citenamefont {Barbato}\ and\ \citenamefont
  {Gautier}(2001)}]{BARBATO2001}%
  \BibitemOpen
  \bibfield  {author} {\bibinfo {author} {\bibfnamefont {S.}~\bibnamefont
  {Barbato}}\ and\ \bibinfo {author} {\bibfnamefont {J.L.}\ \bibnamefont
  {Gautier}},\ }\bibfield  {title} {\enquote {\bibinfo {title} {Hollandite
  cathodes for lithium ion batteries. 2. {Thermodynamic} and kinetics studies
  of lithium insertion into {Ba}{M}{Mn}$_{7}${O}$_{16}$ ({M}={Mg}, {Mn}, {Fe},
  {Ni})},}\ }\href@noop {} {\bibfield  {journal} {\bibinfo  {journal}
  {Electrochim. Acta}\ }\textbf {\bibinfo {volume} {46}},\ \bibinfo {pages}
  {2767 -- 2776} (\bibinfo {year} {2001})}\BibitemShut {NoStop}%
\bibitem [{\citenamefont {Kijima}\ \emph {et~al.}(2005)\citenamefont {Kijima},
  \citenamefont {Takahashi}, \citenamefont {Akimoto},\ and\ \citenamefont
  {Awaka}}]{Kijima2005}%
  \BibitemOpen
  \bibfield  {author} {\bibinfo {author} {\bibfnamefont {N.}~\bibnamefont
  {Kijima}}, \bibinfo {author} {\bibfnamefont {Y.}~\bibnamefont {Takahashi}},
  \bibinfo {author} {\bibfnamefont {J.}~\bibnamefont {Akimoto}}, \ and\
  \bibinfo {author} {\bibfnamefont {J.}~\bibnamefont {Awaka}},\ }\bibfield
  {title} {\enquote {\bibinfo {title} {Lithium ion insertion and extraction
  reactions with hollandite-type manganese dioxide free from any stabilizing
  cations in its tunnel cavity},}\ }\href@noop {} {\bibfield  {journal}
  {\bibinfo  {journal} {J. Solid State Chem.}\ }\textbf {\bibinfo {volume}
  {178}},\ \bibinfo {pages} {2741 -- 2750} (\bibinfo {year}
  {2005})}\BibitemShut {NoStop}%
\bibitem [{\citenamefont {Johnson}(2007)}]{Johnson2007}%
  \BibitemOpen
  \bibfield  {author} {\bibinfo {author} {\bibfnamefont {C.~S.}\ \bibnamefont
  {Johnson}},\ }\bibfield  {title} {\enquote {\bibinfo {title} {Development and
  utility of manganese oxides as cathodes in lithium batteries},}\ }\href@noop
  {} {\bibfield  {journal} {\bibinfo  {journal} {J. Power Sources}\ }\textbf
  {\bibinfo {volume} {165}},\ \bibinfo {pages} {559 -- 565} (\bibinfo {year}
  {2007})}\BibitemShut {NoStop}%
\bibitem [{\citenamefont {Zhang}\ \emph
  {et~al.}(2012{\natexlab{a}})\citenamefont {Zhang}, \citenamefont {Feng},
  \citenamefont {Zhang}, \citenamefont {Guo}, \citenamefont {Chen},
  \citenamefont {Li},\ and\ \citenamefont {Liu}}]{Zhang12}%
  \BibitemOpen
  \bibfield  {author} {\bibinfo {author} {\bibfnamefont {C.}~\bibnamefont
  {Zhang}}, \bibinfo {author} {\bibfnamefont {C.}~\bibnamefont {Feng}},
  \bibinfo {author} {\bibfnamefont {P.}~\bibnamefont {Zhang}}, \bibinfo
  {author} {\bibfnamefont {Z.}~\bibnamefont {Guo}}, \bibinfo {author}
  {\bibfnamefont {Z.}~\bibnamefont {Chen}}, \bibinfo {author} {\bibfnamefont
  {S.}~\bibnamefont {Li}}, \ and\ \bibinfo {author} {\bibfnamefont
  {H.}~\bibnamefont {Liu}},\ }\bibfield  {title} {\enquote {\bibinfo {title}
  {{K}$_{0.25}${Mn}$_2${O}$_4$ nanofiber microclusters as high power cathode
  materials for rechargeable lithium batteries},}\ }\href@noop {} {\bibfield
  {journal} {\bibinfo  {journal} {RSC Adv.}\ }\textbf {\bibinfo {volume} {2}},\
  \bibinfo {pages} {1643--1649} (\bibinfo {year}
  {2012}{\natexlab{a}})}\BibitemShut {NoStop}%
\bibitem [{\citenamefont {Ling}\ and\ \citenamefont {Mizuno}(2012)}]{Ling12}%
  \BibitemOpen
  \bibfield  {author} {\bibinfo {author} {\bibfnamefont {C.}~\bibnamefont
  {Ling}}\ and\ \bibinfo {author} {\bibfnamefont {F.}~\bibnamefont {Mizuno}},\
  }\bibfield  {title} {\enquote {\bibinfo {title} {Capture lithium in
  $\ensuremath{\alpha}$-{Mn}{O}${}_{2}$: Insights from first principles},}\
  }\href@noop {} {\bibfield  {journal} {\bibinfo  {journal} {Chem. Mater.}\
  }\textbf {\bibinfo {volume} {24}},\ \bibinfo {pages} {3943--3951} (\bibinfo
  {year} {2012})}\BibitemShut {NoStop}%
\bibitem [{\citenamefont {Trahey}\ \emph {et~al.}(2013)\citenamefont {Trahey},
  \citenamefont {Karan}, \citenamefont {Chan}, \citenamefont {Lu},
  \citenamefont {Ren}, \citenamefont {Greeley}, \citenamefont
  {Balasubramanian}, \citenamefont {Burrell}, \citenamefont {Curtiss},\ and\
  \citenamefont {Thackeray}}]{Trahey13}%
  \BibitemOpen
  \bibfield  {author} {\bibinfo {author} {\bibfnamefont {L.}~\bibnamefont
  {Trahey}}, \bibinfo {author} {\bibfnamefont {N.~K.}\ \bibnamefont {Karan}},
  \bibinfo {author} {\bibfnamefont {M.~K.~Y.}\ \bibnamefont {Chan}}, \bibinfo
  {author} {\bibfnamefont {J.}~\bibnamefont {Lu}}, \bibinfo {author}
  {\bibfnamefont {Y.}~\bibnamefont {Ren}}, \bibinfo {author} {\bibfnamefont
  {J.}~\bibnamefont {Greeley}}, \bibinfo {author} {\bibfnamefont
  {M.}~\bibnamefont {Balasubramanian}}, \bibinfo {author} {\bibfnamefont
  {A.~K.}\ \bibnamefont {Burrell}}, \bibinfo {author} {\bibfnamefont {L.~A.}\
  \bibnamefont {Curtiss}}, \ and\ \bibinfo {author} {\bibfnamefont {M.~M.}\
  \bibnamefont {Thackeray}},\ }\bibfield  {title} {\enquote {\bibinfo {title}
  {Synthesis, characterization, and structural modeling of high-capacity, dual
  functioning {Mn}{O}$_{2}$ electrode/electrocatalysts for {Li}-{O}$_{2}$
  cells},}\ }\href@noop {} {\bibfield  {journal} {\bibinfo  {journal} {Adv.
  Energy Mater.}\ }\textbf {\bibinfo {volume} {3}},\ \bibinfo {pages} {75--84}
  (\bibinfo {year} {2013})}\BibitemShut {NoStop}%
\bibitem [{\citenamefont {Tompsett}\ and\ \citenamefont
  {Islam}(2013)}]{Islam13}%
  \BibitemOpen
  \bibfield  {author} {\bibinfo {author} {\bibfnamefont {D.~A.}\ \bibnamefont
  {Tompsett}}\ and\ \bibinfo {author} {\bibfnamefont {M.~S.}\ \bibnamefont
  {Islam}},\ }\bibfield  {title} {\enquote {\bibinfo {title} {Electrochemistry
  of hollandite $\ensuremath{\alpha}$-{Mn}{O}$_{2}$: Li-ion and {Na}-ion
  insertion and {Li}$_{2}${O} incorporation},}\ }\href@noop {} {\bibfield
  {journal} {\bibinfo  {journal} {Chem. Mater.}\ }\textbf {\bibinfo {volume}
  {25}},\ \bibinfo {pages} {2515--2526} (\bibinfo {year} {2013})}\BibitemShut
  {NoStop}%
\bibitem [{\citenamefont {Yuan}\ \emph {et~al.}(2015)\citenamefont {Yuan},
  \citenamefont {Nie}, \citenamefont {Odegard}, \citenamefont {Xu},
  \citenamefont {Zhou}, \citenamefont {Santhanagopalan}, \citenamefont {He},
  \citenamefont {Asayesh-Ardakani}, \citenamefont {Meng}, \citenamefont {Klie},
  \citenamefont {Johnson}, \citenamefont {Lu},\ and\ \citenamefont
  {Shahbazian-Yassar}}]{Yuan15}%
  \BibitemOpen
  \bibfield  {author} {\bibinfo {author} {\bibfnamefont {Y.}~\bibnamefont
  {Yuan}}, \bibinfo {author} {\bibfnamefont {A.}~\bibnamefont {Nie}}, \bibinfo
  {author} {\bibfnamefont {G.~M.}\ \bibnamefont {Odegard}}, \bibinfo {author}
  {\bibfnamefont {R.}~\bibnamefont {Xu}}, \bibinfo {author} {\bibfnamefont
  {D.}~\bibnamefont {Zhou}}, \bibinfo {author} {\bibfnamefont {S.}~\bibnamefont
  {Santhanagopalan}}, \bibinfo {author} {\bibfnamefont {K.}~\bibnamefont {He}},
  \bibinfo {author} {\bibfnamefont {H.}~\bibnamefont {Asayesh-Ardakani}},
  \bibinfo {author} {\bibfnamefont {D.~D.}\ \bibnamefont {Meng}}, \bibinfo
  {author} {\bibfnamefont {R.~F.}\ \bibnamefont {Klie}}, \bibinfo {author}
  {\bibfnamefont {C.}~\bibnamefont {Johnson}}, \bibinfo {author} {\bibfnamefont
  {J.}~\bibnamefont {Lu}}, \ and\ \bibinfo {author} {\bibfnamefont
  {R.}~\bibnamefont {Shahbazian-Yassar}},\ }\bibfield  {title} {\enquote
  {\bibinfo {title} {Asynchronous crystal cell expansion during lithiation of
  {K}$^{+}$-stabilized $\ensuremath{\alpha}$-{Mn}{O}$_{2}$},}\ }\href@noop {}
  {\bibfield  {journal} {\bibinfo  {journal} {Nano Lett.}\ }\textbf {\bibinfo
  {volume} {15}},\ \bibinfo {pages} {2998--3007} (\bibinfo {year}
  {2015})}\BibitemShut {NoStop}%
\bibitem [{\citenamefont {Yuan}\ \emph {et~al.}(2016)\citenamefont {Yuan},
  \citenamefont {Zhan}, \citenamefont {He}, \citenamefont {Chen}, \citenamefont
  {Yao}, \citenamefont {Sharifi-Asl}, \citenamefont {Song}, \citenamefont
  {Yang}, \citenamefont {Nie}, \citenamefont {Luo}, \citenamefont {Wang},
  \citenamefont {Wood}, \citenamefont {Amine}, \citenamefont {Islam},
  \citenamefont {Lu},\ and\ \citenamefont {Shahbazian-Yassar}}]{Yuan16}%
  \BibitemOpen
  \bibfield  {author} {\bibinfo {author} {\bibfnamefont {Y.~F.}\ \bibnamefont
  {Yuan}}, \bibinfo {author} {\bibfnamefont {C.}~\bibnamefont {Zhan}}, \bibinfo
  {author} {\bibfnamefont {K.}~\bibnamefont {He}}, \bibinfo {author}
  {\bibfnamefont {H.~R.}\ \bibnamefont {Chen}}, \bibinfo {author}
  {\bibfnamefont {W.~T.}\ \bibnamefont {Yao}}, \bibinfo {author} {\bibfnamefont
  {S.}~\bibnamefont {Sharifi-Asl}}, \bibinfo {author} {\bibfnamefont
  {B.}~\bibnamefont {Song}}, \bibinfo {author} {\bibfnamefont {Z.~Z.}\
  \bibnamefont {Yang}}, \bibinfo {author} {\bibfnamefont {A.~M.}\ \bibnamefont
  {Nie}}, \bibinfo {author} {\bibfnamefont {X.~Y.}\ \bibnamefont {Luo}},
  \bibinfo {author} {\bibfnamefont {H.}~\bibnamefont {Wang}}, \bibinfo {author}
  {\bibfnamefont {S.~M.}\ \bibnamefont {Wood}}, \bibinfo {author}
  {\bibfnamefont {K.}~\bibnamefont {Amine}}, \bibinfo {author} {\bibfnamefont
  {M.~S.}\ \bibnamefont {Islam}}, \bibinfo {author} {\bibfnamefont
  {J.}~\bibnamefont {Lu}}, \ and\ \bibinfo {author} {\bibfnamefont
  {R.}~\bibnamefont {Shahbazian-Yassar}},\ }\bibfield  {title} {\enquote
  {\bibinfo {title} {The influence of large cations on the electrochemical
  properties of tunnel-structured metal oxides},}\ }\href@noop {} {\bibfield
  {journal} {\bibinfo  {journal} {Nat. Commun.}\ }\textbf {\bibinfo {volume}
  {7}},\ \bibinfo {pages} {13374} (\bibinfo {year} {2016})}\BibitemShut
  {NoStop}%
\bibitem [{\citenamefont {Xu}\ \emph {et~al.}(2017)\citenamefont {Xu},
  \citenamefont {Wu}, \citenamefont {Meng}, \citenamefont {Kaltak},
  \citenamefont {Huang}, \citenamefont {Durham}, \citenamefont
  {Fernandez-Serra}, \citenamefont {Sun}, \citenamefont {Marschilok},
  \citenamefont {Takeuchi}, \citenamefont {Takeuchi}, \citenamefont
  {Hybertsen},\ and\ \citenamefont {Zhu}}]{Xu17}%
  \BibitemOpen
  \bibfield  {author} {\bibinfo {author} {\bibfnamefont {F.}~\bibnamefont
  {Xu}}, \bibinfo {author} {\bibfnamefont {L.~J.}\ \bibnamefont {Wu}}, \bibinfo
  {author} {\bibfnamefont {Q.~P.}\ \bibnamefont {Meng}}, \bibinfo {author}
  {\bibfnamefont {M.}~\bibnamefont {Kaltak}}, \bibinfo {author} {\bibfnamefont
  {J.~P.}\ \bibnamefont {Huang}}, \bibinfo {author} {\bibfnamefont {J.~L.}\
  \bibnamefont {Durham}}, \bibinfo {author} {\bibfnamefont {M.}~\bibnamefont
  {Fernandez-Serra}}, \bibinfo {author} {\bibfnamefont {L.~T.}\ \bibnamefont
  {Sun}}, \bibinfo {author} {\bibfnamefont {A.~C.}\ \bibnamefont {Marschilok}},
  \bibinfo {author} {\bibfnamefont {E.~S.}\ \bibnamefont {Takeuchi}}, \bibinfo
  {author} {\bibfnamefont {K.~J.}\ \bibnamefont {Takeuchi}}, \bibinfo {author}
  {\bibfnamefont {M.~S.}\ \bibnamefont {Hybertsen}}, \ and\ \bibinfo {author}
  {\bibfnamefont {Y.~M.}\ \bibnamefont {Zhu}},\ }\bibfield  {title} {\enquote
  {\bibinfo {title} {Visualization of lithium-ion transport and phase evolution
  within and between manganese oxide nanorods},}\ }\href@noop {} {\bibfield
  {journal} {\bibinfo  {journal} {Nature Commun.}\ }\textbf {\bibinfo {volume}
  {8}},\ \bibinfo {pages} {15400} (\bibinfo {year} {2017})}\BibitemShut
  {NoStop}%
\bibitem [{\citenamefont {Yang}\ \emph {et~al.}(2017)\citenamefont {Yang},
  \citenamefont {Ford}, \citenamefont {Park}, \citenamefont {Ren},
  \citenamefont {Kim}, \citenamefont {Kim}, \citenamefont {Fister},
  \citenamefont {Chan},\ and\ \citenamefont {Thackeray}}]{Yang17}%
  \BibitemOpen
  \bibfield  {author} {\bibinfo {author} {\bibfnamefont {Z.~Z.}\ \bibnamefont
  {Yang}}, \bibinfo {author} {\bibfnamefont {D.~C.}\ \bibnamefont {Ford}},
  \bibinfo {author} {\bibfnamefont {J.~S.}\ \bibnamefont {Park}}, \bibinfo
  {author} {\bibfnamefont {Y.}~\bibnamefont {Ren}}, \bibinfo {author}
  {\bibfnamefont {S.}~\bibnamefont {Kim}}, \bibinfo {author} {\bibfnamefont
  {H.}~\bibnamefont {Kim}}, \bibinfo {author} {\bibfnamefont {T.~T.}\
  \bibnamefont {Fister}}, \bibinfo {author} {\bibfnamefont {M.~K.~Y.}\
  \bibnamefont {Chan}}, \ and\ \bibinfo {author} {\bibfnamefont {M.~M.}\
  \bibnamefont {Thackeray}},\ }\bibfield  {title} {\enquote {\bibinfo {title}
  {Probing the release and uptake of water in
  $\alpha$-{Mn}{O}$_{2}\cdot${x}{H}$_{2}${O}},}\ }\href@noop {} {\bibfield
  {journal} {\bibinfo  {journal} {Chem. Mater.}\ }\textbf {\bibinfo {volume}
  {29}},\ \bibinfo {pages} {1507--1517} (\bibinfo {year} {2017})}\BibitemShut
  {NoStop}%
\bibitem [{\citenamefont {Doeff}\ \emph {et~al.}(1994)\citenamefont {Doeff},
  \citenamefont {Peng}, \citenamefont {Ma},\ and\ \citenamefont
  {De~Jonghe}}]{Doeff94}%
  \BibitemOpen
  \bibfield  {author} {\bibinfo {author} {\bibfnamefont {M.~M.}\ \bibnamefont
  {Doeff}}, \bibinfo {author} {\bibfnamefont {M.~Y.}\ \bibnamefont {Peng}},
  \bibinfo {author} {\bibfnamefont {Y.}~\bibnamefont {Ma}}, \ and\ \bibinfo
  {author} {\bibfnamefont {L.~C.}\ \bibnamefont {De~Jonghe}},\ }\bibfield
  {title} {\enquote {\bibinfo {title} {Orthorhombic {Na}$_x${Mn}{O}$_2$ as a
  cathode material for secondary sodium and lithium polymer batteries},}\
  }\href@noop {} {\bibfield  {journal} {\bibinfo  {journal} {J. Electrochem.
  Soc.}\ }\textbf {\bibinfo {volume} {141}},\ \bibinfo {pages} {L145--L147}
  (\bibinfo {year} {1994})}\BibitemShut {NoStop}%
\bibitem [{\citenamefont {Sauvage}\ \emph {et~al.}(2007)\citenamefont
  {Sauvage}, \citenamefont {Laffont}, \citenamefont {Tarascon},\ and\
  \citenamefont {Baudrin}}]{Sauvage07}%
  \BibitemOpen
  \bibfield  {author} {\bibinfo {author} {\bibfnamefont {F.}~\bibnamefont
  {Sauvage}}, \bibinfo {author} {\bibfnamefont {L.}~\bibnamefont {Laffont}},
  \bibinfo {author} {\bibfnamefont {J.-M.}\ \bibnamefont {Tarascon}}, \ and\
  \bibinfo {author} {\bibfnamefont {E.}~\bibnamefont {Baudrin}},\ }\bibfield
  {title} {\enquote {\bibinfo {title} {Study of the insertion/deinsertion
  mechanism of sodium into {Na}$_{0.44}${Mn}{O}$_{2}$},}\ }\href@noop {}
  {\bibfield  {journal} {\bibinfo  {journal} {Inorg. Chem.}\ }\textbf {\bibinfo
  {volume} {46}},\ \bibinfo {pages} {3289--3294} (\bibinfo {year}
  {2007})}\BibitemShut {NoStop}%
\bibitem [{\citenamefont {Kim}\ \emph {et~al.}(2012)\citenamefont {Kim},
  \citenamefont {Kim}, \citenamefont {Seo}, \citenamefont {Yeom}, \citenamefont
  {Kang}, \citenamefont {Kim},\ and\ \citenamefont {Jung}}]{Kim12}%
  \BibitemOpen
  \bibfield  {author} {\bibinfo {author} {\bibfnamefont {H.}~\bibnamefont
  {Kim}}, \bibinfo {author} {\bibfnamefont {D.~J.}\ \bibnamefont {Kim}},
  \bibinfo {author} {\bibfnamefont {D.-H.}\ \bibnamefont {Seo}}, \bibinfo
  {author} {\bibfnamefont {M.~S.}\ \bibnamefont {Yeom}}, \bibinfo {author}
  {\bibfnamefont {K.}~\bibnamefont {Kang}}, \bibinfo {author} {\bibfnamefont
  {D.~K.}\ \bibnamefont {Kim}}, \ and\ \bibinfo {author} {\bibfnamefont
  {Y.}~\bibnamefont {Jung}},\ }\bibfield  {title} {\enquote {\bibinfo {title}
  {Ab initio study of the sodium intercalation and intermediate phases in
  {Na}$_{0.44}${Mn}{O}$_{2}$ for sodium-ion battery},}\ }\href@noop {}
  {\bibfield  {journal} {\bibinfo  {journal} {Chem. Mater.}\ }\textbf {\bibinfo
  {volume} {24}},\ \bibinfo {pages} {1205--1211} (\bibinfo {year}
  {2012})}\BibitemShut {NoStop}%
\bibitem [{\citenamefont {Perez-Flores}\ \emph {et~al.}(2014)\citenamefont
  {Perez-Flores}, \citenamefont {Baehtz}, \citenamefont {Kuhn},\ and\
  \citenamefont {Garcia-Alvarado}}]{PerezFlores14}%
  \BibitemOpen
  \bibfield  {author} {\bibinfo {author} {\bibfnamefont {J.~C.}\ \bibnamefont
  {Perez-Flores}}, \bibinfo {author} {\bibfnamefont {C.}~\bibnamefont
  {Baehtz}}, \bibinfo {author} {\bibfnamefont {A.}~\bibnamefont {Kuhn}}, \ and\
  \bibinfo {author} {\bibfnamefont {F.}~\bibnamefont {Garcia-Alvarado}},\
  }\bibfield  {title} {\enquote {\bibinfo {title} {Hollandite-type
  {Ti}{O}$_{2}$: a new negative electrode material for sodium-ion batteries},}\
  }\href@noop {} {\bibfield  {journal} {\bibinfo  {journal} {J. Mater. Chem.
  A}\ }\textbf {\bibinfo {volume} {2}},\ \bibinfo {pages} {1825--1833}
  (\bibinfo {year} {2014})}\BibitemShut {NoStop}%
\bibitem [{\citenamefont {Huang}\ \emph {et~al.}(2017)\citenamefont {Huang},
  \citenamefont {Poyraz}, \citenamefont {Lee}, \citenamefont {Wu},
  \citenamefont {Zhu}, \citenamefont {Marschilok}, \citenamefont {Takeuchi},\
  and\ \citenamefont {Takeuchi}}]{Huang17}%
  \BibitemOpen
  \bibfield  {author} {\bibinfo {author} {\bibfnamefont {J.~P.}\ \bibnamefont
  {Huang}}, \bibinfo {author} {\bibfnamefont {A.~S.}\ \bibnamefont {Poyraz}},
  \bibinfo {author} {\bibfnamefont {S.~Y.}\ \bibnamefont {Lee}}, \bibinfo
  {author} {\bibfnamefont {L.~J.}\ \bibnamefont {Wu}}, \bibinfo {author}
  {\bibfnamefont {Y.~M.}\ \bibnamefont {Zhu}}, \bibinfo {author} {\bibfnamefont
  {A.~C.}\ \bibnamefont {Marschilok}}, \bibinfo {author} {\bibfnamefont
  {K.~J.}\ \bibnamefont {Takeuchi}}, \ and\ \bibinfo {author} {\bibfnamefont
  {E.~S.}\ \bibnamefont {Takeuchi}},\ }\bibfield  {title} {\enquote {\bibinfo
  {title} {Silver-containing $\alpha$-{Mn}{O}$_{2}$ nanorods: Electrochemistry
  in {Na}-based battery systems},}\ }\href@noop {} {\bibfield  {journal}
  {\bibinfo  {journal} {{ACS} Appl. Mater. Interfaces}\ }\textbf {\bibinfo
  {volume} {9}},\ \bibinfo {pages} {4333--4342} (\bibinfo {year}
  {2017})}\BibitemShut {NoStop}%
\bibitem [{\citenamefont {Zhang}\ \emph
  {et~al.}(2012{\natexlab{b}})\citenamefont {Zhang}, \citenamefont {Yu},
  \citenamefont {Nam}, \citenamefont {Ling}, \citenamefont {Arthur},
  \citenamefont {Song}, \citenamefont {Knapp}, \citenamefont {Ehrlich},
  \citenamefont {Yang},\ and\ \citenamefont {Matsui}}]{Zhang2012}%
  \BibitemOpen
  \bibfield  {author} {\bibinfo {author} {\bibfnamefont {R.}~\bibnamefont
  {Zhang}}, \bibinfo {author} {\bibfnamefont {X.}~\bibnamefont {Yu}}, \bibinfo
  {author} {\bibfnamefont {K.-W.}\ \bibnamefont {Nam}}, \bibinfo {author}
  {\bibfnamefont {C.}~\bibnamefont {Ling}}, \bibinfo {author} {\bibfnamefont
  {T.~S.}\ \bibnamefont {Arthur}}, \bibinfo {author} {\bibfnamefont
  {W.}~\bibnamefont {Song}}, \bibinfo {author} {\bibfnamefont {A.~M.}\
  \bibnamefont {Knapp}}, \bibinfo {author} {\bibfnamefont {S.~N.}\ \bibnamefont
  {Ehrlich}}, \bibinfo {author} {\bibfnamefont {X.-Q.}\ \bibnamefont {Yang}}, \
  and\ \bibinfo {author} {\bibfnamefont {M.}~\bibnamefont {Matsui}},\
  }\bibfield  {title} {\enquote {\bibinfo {title} {$\alpha$-{Mn}{O}$_2$ as a
  cathode material for rechargeable {Mg} batteries},}\ }\href@noop {}
  {\bibfield  {journal} {\bibinfo  {journal} {Electrochem. Commun.}\ }\textbf
  {\bibinfo {volume} {23}},\ \bibinfo {pages} {110 -- 113} (\bibinfo {year}
  {2012}{\natexlab{b}})}\BibitemShut {NoStop}%
\bibitem [{\citenamefont {Huang}\ \emph {et~al.}(2016)\citenamefont {Huang},
  \citenamefont {Poyraz}, \citenamefont {Takeuchi}, \citenamefont {Takeuchi},\
  and\ \citenamefont {Marschilok}}]{Huang16}%
  \BibitemOpen
  \bibfield  {author} {\bibinfo {author} {\bibfnamefont {J.~P.}\ \bibnamefont
  {Huang}}, \bibinfo {author} {\bibfnamefont {A.~S.}\ \bibnamefont {Poyraz}},
  \bibinfo {author} {\bibfnamefont {K.~J.}\ \bibnamefont {Takeuchi}}, \bibinfo
  {author} {\bibfnamefont {E.~S.}\ \bibnamefont {Takeuchi}}, \ and\ \bibinfo
  {author} {\bibfnamefont {A.~C.}\ \bibnamefont {Marschilok}},\ }\bibfield
  {title} {\enquote {\bibinfo {title} {{M}$_{x}${Mn}$_{8}${O}$_{16}$ ({M} =
  {Ag} or {K}) as promising cathode materials for secondary {Mg} based
  batteries: The role of the cation {M}},}\ }\href@noop {} {\bibfield
  {journal} {\bibinfo  {journal} {Chem. Commun.}\ }\textbf {\bibinfo {volume}
  {52}},\ \bibinfo {pages} {4088--4091} (\bibinfo {year} {2016})}\BibitemShut
  {NoStop}%
\bibitem [{\citenamefont {Zhu}\ \emph {et~al.}(2010)\citenamefont {Zhu},
  \citenamefont {Marschilok}, \citenamefont {Lee}, \citenamefont {Takeuchi},\
  and\ \citenamefont {Takeuchi}}]{Zhu2010}%
  \BibitemOpen
  \bibfield  {author} {\bibinfo {author} {\bibfnamefont {S.}~\bibnamefont
  {Zhu}}, \bibinfo {author} {\bibfnamefont {A.~C.}\ \bibnamefont {Marschilok}},
  \bibinfo {author} {\bibfnamefont {C.-Y.}\ \bibnamefont {Lee}}, \bibinfo
  {author} {\bibfnamefont {E.~S.}\ \bibnamefont {Takeuchi}}, \ and\ \bibinfo
  {author} {\bibfnamefont {K.~J.}\ \bibnamefont {Takeuchi}},\ }\bibfield
  {title} {\enquote {\bibinfo {title} {Synthesis and electrochemistry of silver
  hollandite},}\ }\href@noop {} {\bibfield  {journal} {\bibinfo  {journal}
  {Electrochem. Solid State Lett.}\ }\textbf {\bibinfo {volume} {13}},\
  \bibinfo {pages} {A98--A100} (\bibinfo {year} {2010})}\BibitemShut {NoStop}%
\bibitem [{\citenamefont {Takeuchi}\ \emph {et~al.}(2012)\citenamefont
  {Takeuchi}, \citenamefont {Yau}, \citenamefont {Menard}, \citenamefont
  {Marschilok},\ and\ \citenamefont {Takeuchi}}]{Takeuchi12}%
  \BibitemOpen
  \bibfield  {author} {\bibinfo {author} {\bibfnamefont {K.~J.}\ \bibnamefont
  {Takeuchi}}, \bibinfo {author} {\bibfnamefont {S.~Z.}\ \bibnamefont {Yau}},
  \bibinfo {author} {\bibfnamefont {M.~C.}\ \bibnamefont {Menard}}, \bibinfo
  {author} {\bibfnamefont {A.~C.}\ \bibnamefont {Marschilok}}, \ and\ \bibinfo
  {author} {\bibfnamefont {E.~S.}\ \bibnamefont {Takeuchi}},\ }\bibfield
  {title} {\enquote {\bibinfo {title} {Synthetic control of composition and
  crystallite size of silver hollandite, {Ag}$_x${Mn}$_{8}${O}$_{16}$: Impact
  on electrochemistry},}\ }\href@noop {} {\bibfield  {journal} {\bibinfo
  {journal} {ACS Appl. Mater. Interfaces}\ }\textbf {\bibinfo {volume} {4}},\
  \bibinfo {pages} {5547--5554} (\bibinfo {year} {2012})}\BibitemShut {NoStop}%
\bibitem [{\citenamefont {Takeuchi}\ \emph {et~al.}(2013)\citenamefont
  {Takeuchi}, \citenamefont {Yau}, \citenamefont {Subramanian}, \citenamefont
  {Marschilok},\ and\ \citenamefont {Takeuchi}}]{Takeuchi13}%
  \BibitemOpen
  \bibfield  {author} {\bibinfo {author} {\bibfnamefont {K.~J.}\ \bibnamefont
  {Takeuchi}}, \bibinfo {author} {\bibfnamefont {S.~Z.}\ \bibnamefont {Yau}},
  \bibinfo {author} {\bibfnamefont {A.}~\bibnamefont {Subramanian}}, \bibinfo
  {author} {\bibfnamefont {A.~C.}\ \bibnamefont {Marschilok}}, \ and\ \bibinfo
  {author} {\bibfnamefont {E.~S.}\ \bibnamefont {Takeuchi}},\ }\bibfield
  {title} {\enquote {\bibinfo {title} {The electrochemistry of silver
  hollandite nanorods, {Ag}$_{x}${Mn}$_{8}${O}$_{16}$: Enhancement of
  electrochemical battery performance via dimensional and compositional
  control},}\ }\href@noop {} {\bibfield  {journal} {\bibinfo  {journal} {J.
  Electrochem. Soc.}\ }\textbf {\bibinfo {volume} {160}},\ \bibinfo {pages}
  {A3090--A3094} (\bibinfo {year} {2013})}\BibitemShut {NoStop}%
\bibitem [{\citenamefont {Wu}\ \emph {et~al.}(2015)\citenamefont {Wu},
  \citenamefont {Xu}, \citenamefont {Zhu}, \citenamefont {Brady}, \citenamefont
  {Huang}, \citenamefont {Durham}, \citenamefont {Dooryhee}, \citenamefont
  {Marschilok}, \citenamefont {Takeuchi},\ and\ \citenamefont
  {J.Takeuchi}}]{Wu15}%
  \BibitemOpen
  \bibfield  {author} {\bibinfo {author} {\bibfnamefont {L.}~\bibnamefont
  {Wu}}, \bibinfo {author} {\bibfnamefont {F.}~\bibnamefont {Xu}}, \bibinfo
  {author} {\bibfnamefont {Y.}~\bibnamefont {Zhu}}, \bibinfo {author}
  {\bibfnamefont {A.~B.}\ \bibnamefont {Brady}}, \bibinfo {author}
  {\bibfnamefont {J.}~\bibnamefont {Huang}}, \bibinfo {author} {\bibfnamefont
  {J.~L.}\ \bibnamefont {Durham}}, \bibinfo {author} {\bibfnamefont
  {E.}~\bibnamefont {Dooryhee}}, \bibinfo {author} {\bibfnamefont {A.~C.}\
  \bibnamefont {Marschilok}}, \bibinfo {author} {\bibfnamefont {E.~S.}\
  \bibnamefont {Takeuchi}}, \ and\ \bibinfo {author} {\bibfnamefont
  {K.}~\bibnamefont {J.Takeuchi}},\ }\bibfield  {title} {\enquote {\bibinfo
  {title} {Structural defects of silver hollandite,
  {Ag}$_{x}${Mn}$_{8}${O}$_{y}$, nanorods: Dramatic impact on
  electrochemistry},}\ }\href@noop {} {\bibfield  {journal} {\bibinfo
  {journal} {ACS Nano}\ }\textbf {\bibinfo {volume} {9}},\ \bibinfo {pages}
  {8430--8439} (\bibinfo {year} {2015})}\BibitemShut {NoStop}%
\bibitem [{\citenamefont {Momma}\ and\ \citenamefont {Izumi}(2011)}]{VESTA}%
  \BibitemOpen
  \bibfield  {author} {\bibinfo {author} {\bibfnamefont {K.}~\bibnamefont
  {Momma}}\ and\ \bibinfo {author} {\bibfnamefont {F.}~\bibnamefont {Izumi}},\
  }\bibfield  {title} {\enquote {\bibinfo {title} {{{\it VESTA3} for
  three-dimensional visualization of crystal, volumetric and morphology
  data}},}\ }\href@noop {} {\bibfield  {journal} {\bibinfo  {journal} {J. Appl.
  Cryst.}\ }\textbf {\bibinfo {volume} {44}},\ \bibinfo {pages} {1272--1276}
  (\bibinfo {year} {2011})}\BibitemShut {NoStop}%
\bibitem [{\citenamefont {Kijima}\ \emph {et~al.}(2004)\citenamefont {Kijima},
  \citenamefont {Ikeda}, \citenamefont {Oikawa}, \citenamefont {Izumi},\ and\
  \citenamefont {Yoshimura}}]{Kijima2004}%
  \BibitemOpen
  \bibfield  {author} {\bibinfo {author} {\bibfnamefont {N.}~\bibnamefont
  {Kijima}}, \bibinfo {author} {\bibfnamefont {T.}~\bibnamefont {Ikeda}},
  \bibinfo {author} {\bibfnamefont {K.}~\bibnamefont {Oikawa}}, \bibinfo
  {author} {\bibfnamefont {F.}~\bibnamefont {Izumi}}, \ and\ \bibinfo {author}
  {\bibfnamefont {Y.}~\bibnamefont {Yoshimura}},\ }\bibfield  {title} {\enquote
  {\bibinfo {title} {Crystal structure of an open-tunnel oxide
  $\alpha$-{Mn}{O}$_{2}$ analyzed by {Rietveld} refinements and {MEM}-based
  pattern fitting},}\ }\href@noop {} {\bibfield  {journal} {\bibinfo  {journal}
  {J. Solid State Chem.}\ }\textbf {\bibinfo {volume} {177}},\ \bibinfo {pages}
  {1258 -- 1267} (\bibinfo {year} {2004})}\BibitemShut {NoStop}%
\bibitem [{\citenamefont {Isobe}\ \emph {et~al.}(2006)\citenamefont {Isobe},
  \citenamefont {Koishi}, \citenamefont {Kouno}, \citenamefont {Yamaura},
  \citenamefont {Yamauchi}, \citenamefont {Ueda}, \citenamefont {Gotou},
  \citenamefont {Yagi},\ and\ \citenamefont {Ueda}}]{Isobe06}%
  \BibitemOpen
  \bibfield  {author} {\bibinfo {author} {\bibfnamefont {M.}~\bibnamefont
  {Isobe}}, \bibinfo {author} {\bibfnamefont {S.}~\bibnamefont {Koishi}},
  \bibinfo {author} {\bibfnamefont {N.}~\bibnamefont {Kouno}}, \bibinfo
  {author} {\bibfnamefont {J.-I.}\ \bibnamefont {Yamaura}}, \bibinfo {author}
  {\bibfnamefont {T.}~\bibnamefont {Yamauchi}}, \bibinfo {author}
  {\bibfnamefont {H.}~\bibnamefont {Ueda}}, \bibinfo {author} {\bibfnamefont
  {H.}~\bibnamefont {Gotou}}, \bibinfo {author} {\bibfnamefont
  {T.}~\bibnamefont {Yagi}}, \ and\ \bibinfo {author} {\bibfnamefont
  {Y.}~\bibnamefont {Ueda}},\ }\bibfield  {title} {\enquote {\bibinfo {title}
  {Observation of metal-insulator transition in hollandite vanadate,
  {K}$_{2}${V}$_{8}${O}$_{16}$},}\ }\href@noop {} {\bibfield  {journal}
  {\bibinfo  {journal} {J. Phys. Soc. Jpn.}\ }\textbf {\bibinfo {volume}
  {75}},\ \bibinfo {pages} {073801} (\bibinfo {year} {2006})}\BibitemShut
  {NoStop}%
\bibitem [{\citenamefont {Komarek}\ \emph {et~al.}(2011)\citenamefont
  {Komarek}, \citenamefont {Isobe}, \citenamefont {Hemberger}, \citenamefont
  {Meier}, \citenamefont {Lorenz}, \citenamefont {Trots}, \citenamefont
  {Cervellino}, \citenamefont {Fernández-Díaz}, \citenamefont {Ueda},\ and\
  \citenamefont {Braden}}]{Komarek11}%
  \BibitemOpen
  \bibfield  {author} {\bibinfo {author} {\bibfnamefont {A.~C.}\ \bibnamefont
  {Komarek}}, \bibinfo {author} {\bibfnamefont {M.}~\bibnamefont {Isobe}},
  \bibinfo {author} {\bibfnamefont {J.}~\bibnamefont {Hemberger}}, \bibinfo
  {author} {\bibfnamefont {D.}~\bibnamefont {Meier}}, \bibinfo {author}
  {\bibfnamefont {T.}~\bibnamefont {Lorenz}}, \bibinfo {author} {\bibfnamefont
  {D.}~\bibnamefont {Trots}}, \bibinfo {author} {\bibfnamefont
  {A.}~\bibnamefont {Cervellino}}, \bibinfo {author} {\bibfnamefont {M.~T.}\
  \bibnamefont {Fernández-Díaz}}, \bibinfo {author} {\bibfnamefont
  {Y.}~\bibnamefont {Ueda}}, \ and\ \bibinfo {author} {\bibfnamefont
  {M.}~\bibnamefont {Braden}},\ }\bibfield  {title} {\enquote {\bibinfo {title}
  {Dimerization and charge order in hollandite {K}$_{2}${V}$_{8}${O}$_{16}$},}\
  }\href@noop {} {\bibfield  {journal} {\bibinfo  {journal} {Phys. Rev. Lett.}\
  }\textbf {\bibinfo {volume} {107}},\ \bibinfo {pages} {027201} (\bibinfo
  {year} {2011})}\BibitemShut {NoStop}%
\bibitem [{\citenamefont {Kim}\ \emph {et~al.}(2016)\citenamefont {Kim},
  \citenamefont {Kim}, \citenamefont {Kim},\ and\ \citenamefont {Min}}]{Kim16}%
  \BibitemOpen
  \bibfield  {author} {\bibinfo {author} {\bibfnamefont {S.}~\bibnamefont
  {Kim}}, \bibinfo {author} {\bibfnamefont {B.~H.}\ \bibnamefont {Kim}},
  \bibinfo {author} {\bibfnamefont {K.}~\bibnamefont {Kim}}, \ and\ \bibinfo
  {author} {\bibfnamefont {B.~I.}\ \bibnamefont {Min}},\ }\bibfield  {title}
  {\enquote {\bibinfo {title} {Metal-insulator transition in a
  spin-orbital-lattice coupled mott system: {K}$_{2}${V}$_{8}${O}$_{16}$},}\
  }\href@noop {} {\bibfield  {journal} {\bibinfo  {journal} {Phys. Rev. B}\
  }\textbf {\bibinfo {volume} {93}},\ \bibinfo {pages} {045106} (\bibinfo
  {year} {2016})}\BibitemShut {NoStop}%
\bibitem [{\citenamefont {Isobe}\ \emph {et~al.}(2009)\citenamefont {Isobe},
  \citenamefont {Koishi}, \citenamefont {Yamazaki}, \citenamefont {Yamaura},
  \citenamefont {Gotou}, \citenamefont {Yagi},\ and\ \citenamefont
  {Ueda}}]{Isobe09}%
  \BibitemOpen
  \bibfield  {author} {\bibinfo {author} {\bibfnamefont {M.}~\bibnamefont
  {Isobe}}, \bibinfo {author} {\bibfnamefont {S.}~\bibnamefont {Koishi}},
  \bibinfo {author} {\bibfnamefont {S.}~\bibnamefont {Yamazaki}}, \bibinfo
  {author} {\bibfnamefont {J.}~\bibnamefont {Yamaura}}, \bibinfo {author}
  {\bibfnamefont {H.}~\bibnamefont {Gotou}}, \bibinfo {author} {\bibfnamefont
  {T.}~\bibnamefont {Yagi}}, \ and\ \bibinfo {author} {\bibfnamefont
  {Y.}~\bibnamefont {Ueda}},\ }\bibfield  {title} {\enquote {\bibinfo {title}
  {Substitution effect on metal-insulator transition of
  {K}$_{2}${V}$_{8}${O}$_{16}$},}\ }\href@noop {} {\bibfield  {journal}
  {\bibinfo  {journal} {J. Phys. Soc. Jpn.}\ }\textbf {\bibinfo {volume} {78}}
  (\bibinfo {year} {2009})}\BibitemShut {NoStop}%
\bibitem [{\citenamefont {Hasegawa}\ \emph {et~al.}(2009)\citenamefont
  {Hasegawa}, \citenamefont {Isobe}, \citenamefont {Yamauchi}, \citenamefont
  {Ueda}, \citenamefont {Yamaura}, \citenamefont {Gotou}, \citenamefont {Yagi},
  \citenamefont {Sato},\ and\ \citenamefont {Ueda}}]{Hasegawa09}%
  \BibitemOpen
  \bibfield  {author} {\bibinfo {author} {\bibfnamefont {K.}~\bibnamefont
  {Hasegawa}}, \bibinfo {author} {\bibfnamefont {M.}~\bibnamefont {Isobe}},
  \bibinfo {author} {\bibfnamefont {T.}~\bibnamefont {Yamauchi}}, \bibinfo
  {author} {\bibfnamefont {H.}~\bibnamefont {Ueda}}, \bibinfo {author}
  {\bibfnamefont {J.-I.}\ \bibnamefont {Yamaura}}, \bibinfo {author}
  {\bibfnamefont {H.}~\bibnamefont {Gotou}}, \bibinfo {author} {\bibfnamefont
  {T.}~\bibnamefont {Yagi}}, \bibinfo {author} {\bibfnamefont {H.}~\bibnamefont
  {Sato}}, \ and\ \bibinfo {author} {\bibfnamefont {Y.}~\bibnamefont {Ueda}},\
  }\bibfield  {title} {\enquote {\bibinfo {title} {Discovery of
  ferromagnetic-half-metal$-$to$-$insulator transition in
  {K}$_{2}${Cr}$_{8}${O}$_{16}$},}\ }\href@noop {} {\bibfield  {journal}
  {\bibinfo  {journal} {Phys. Rev. Lett.}\ }\textbf {\bibinfo {volume} {103}},\
  \bibinfo {pages} {146403} (\bibinfo {year} {2009})}\BibitemShut {NoStop}%
\bibitem [{\citenamefont {Toriyama}\ \emph {et~al.}(2011)\citenamefont
  {Toriyama}, \citenamefont {Nakao}, \citenamefont {Yamaki}, \citenamefont
  {Nakao}, \citenamefont {Murakami}, \citenamefont {Hasegawa}, \citenamefont
  {Isobe}, \citenamefont {Ueda}, \citenamefont {Ushakov}, \citenamefont
  {Khomskii}, \citenamefont {Streltsov}, \citenamefont {Konishi},\ and\
  \citenamefont {Ohta}}]{Toriyama11}%
  \BibitemOpen
  \bibfield  {author} {\bibinfo {author} {\bibfnamefont {T.}~\bibnamefont
  {Toriyama}}, \bibinfo {author} {\bibfnamefont {A.}~\bibnamefont {Nakao}},
  \bibinfo {author} {\bibfnamefont {Y.}~\bibnamefont {Yamaki}}, \bibinfo
  {author} {\bibfnamefont {H.}~\bibnamefont {Nakao}}, \bibinfo {author}
  {\bibfnamefont {Y.}~\bibnamefont {Murakami}}, \bibinfo {author}
  {\bibfnamefont {K.}~\bibnamefont {Hasegawa}}, \bibinfo {author}
  {\bibfnamefont {M.}~\bibnamefont {Isobe}}, \bibinfo {author} {\bibfnamefont
  {Y.}~\bibnamefont {Ueda}}, \bibinfo {author} {\bibfnamefont {A.~V.}\
  \bibnamefont {Ushakov}}, \bibinfo {author} {\bibfnamefont {D.~I.}\
  \bibnamefont {Khomskii}}, \bibinfo {author} {\bibfnamefont {S.~V.}\
  \bibnamefont {Streltsov}}, \bibinfo {author} {\bibfnamefont {T.}~\bibnamefont
  {Konishi}}, \ and\ \bibinfo {author} {\bibfnamefont {Y.}~\bibnamefont
  {Ohta}},\ }\bibfield  {title} {\enquote {\bibinfo {title} {Peierls mechanism
  of the metal-insulator transition in ferromagnetic hollandite
  {K}$_{2}${Cr}$_{8}${O}$_{16}$},}\ }\href@noop {} {\bibfield  {journal}
  {\bibinfo  {journal} {Phys. Rev. Lett.}\ }\textbf {\bibinfo {volume} {107}},\
  \bibinfo {pages} {266402} (\bibinfo {year} {2011})}\BibitemShut {NoStop}%
\bibitem [{\citenamefont {Kim}\ \emph {et~al.}(2014)\citenamefont {Kim},
  \citenamefont {Kim},\ and\ \citenamefont {Min}}]{Kim14}%
  \BibitemOpen
  \bibfield  {author} {\bibinfo {author} {\bibfnamefont {S.}~\bibnamefont
  {Kim}}, \bibinfo {author} {\bibfnamefont {K.}~\bibnamefont {Kim}}, \ and\
  \bibinfo {author} {\bibfnamefont {B.~I.}\ \bibnamefont {Min}},\ }\bibfield
  {title} {\enquote {\bibinfo {title} {Structural instability and the
  mott-peierls transition in a half-metallic hollandite :
  {K}$_{2}${Cr}$_{8}${O}$_{16}$},}\ }\href@noop {} {\bibfield  {journal}
  {\bibinfo  {journal} {Phys. Rev. B}\ }\textbf {\bibinfo {volume} {90}},\
  \bibinfo {pages} {045124} (\bibinfo {year} {2014})}\BibitemShut {NoStop}%
\bibitem [{\citenamefont {Yamamoto}\ \emph {et~al.}(1974)\citenamefont
  {Yamamoto}, \citenamefont {Endo}, \citenamefont {Shimada},\ and\
  \citenamefont {Takada}}]{Yamamoto74}%
  \BibitemOpen
  \bibfield  {author} {\bibinfo {author} {\bibfnamefont {N.}~\bibnamefont
  {Yamamoto}}, \bibinfo {author} {\bibfnamefont {T.}~\bibnamefont {Endo}},
  \bibinfo {author} {\bibfnamefont {M.}~\bibnamefont {Shimada}}, \ and\
  \bibinfo {author} {\bibfnamefont {T.}~\bibnamefont {Takada}},\ }\bibfield
  {title} {\enquote {\bibinfo {title} {Single-crystal growth of
  $\alpha$-{Mn}{O}$_{2}$},}\ }\href@noop {} {\bibfield  {journal} {\bibinfo
  {journal} {Jpn. J. Appl. Phys.}\ }\textbf {\bibinfo {volume} {13}},\ \bibinfo
  {pages} {723--724} (\bibinfo {year} {1974})}\BibitemShut {NoStop}%
\bibitem [{\citenamefont {Strobel}\ \emph {et~al.}(1984)\citenamefont
  {Strobel}, \citenamefont {Vicat},\ and\ \citenamefont {Qui}}]{Strobel84}%
  \BibitemOpen
  \bibfield  {author} {\bibinfo {author} {\bibfnamefont {P.}~\bibnamefont
  {Strobel}}, \bibinfo {author} {\bibfnamefont {J.}~\bibnamefont {Vicat}}, \
  and\ \bibinfo {author} {\bibfnamefont {D.~T.}\ \bibnamefont {Qui}},\
  }\bibfield  {title} {\enquote {\bibinfo {title} {Thermal and
  physical-properties of hollandite-type {K}$_{1.3}${Mn}$_{8}${O}$_{16}$ and
  ({K},{H}$_{3}${O})$_{x}${Mn}$_{8}${O}$_{16}$},}\ }\href@noop {} {\bibfield
  {journal} {\bibinfo  {journal} {J. Solid State Chem.}\ }\textbf {\bibinfo
  {volume} {55}},\ \bibinfo {pages} {67--73} (\bibinfo {year}
  {1984})}\BibitemShut {NoStop}%
\bibitem [{\citenamefont {Sato}\ \emph {et~al.}(1999)\citenamefont {Sato},
  \citenamefont {Enoki}, \citenamefont {Yamaura},\ and\ \citenamefont
  {Yamamoto}}]{Sato99}%
  \BibitemOpen
  \bibfield  {author} {\bibinfo {author} {\bibfnamefont {H.}~\bibnamefont
  {Sato}}, \bibinfo {author} {\bibfnamefont {T.}~\bibnamefont {Enoki}},
  \bibinfo {author} {\bibfnamefont {J.-I.}\ \bibnamefont {Yamaura}}, \ and\
  \bibinfo {author} {\bibfnamefont {N.}~\bibnamefont {Yamamoto}},\ }\bibfield
  {title} {\enquote {\bibinfo {title} {Charge localization and successive
  magnetic phase transitions of mixed-valence manganese oxides
  {K}$_{1.5}${({H}$_{3}${O})}$_{x}${Mn}$_{8}${O}$_{16}$},}\ }\href@noop {}
  {\bibfield  {journal} {\bibinfo  {journal} {Phys. Rev. B}\ }\textbf {\bibinfo
  {volume} {59}},\ \bibinfo {pages} {12836--12841} (\bibinfo {year}
  {1999})}\BibitemShut {NoStop}%
\bibitem [{\citenamefont {Ishiwata}\ \emph {et~al.}(2006)\citenamefont
  {Ishiwata}, \citenamefont {Bos}, \citenamefont {Huang},\ and\ \citenamefont
  {Cava}}]{Ishiwata06}%
  \BibitemOpen
  \bibfield  {author} {\bibinfo {author} {\bibfnamefont {S}~\bibnamefont
  {Ishiwata}}, \bibinfo {author} {\bibfnamefont {J~W~G}\ \bibnamefont {Bos}},
  \bibinfo {author} {\bibfnamefont {Q}~\bibnamefont {Huang}}, \ and\ \bibinfo
  {author} {\bibfnamefont {R~J}\ \bibnamefont {Cava}},\ }\bibfield  {title}
  {\enquote {\bibinfo {title} {Structure and magnetic properties of hollandite
  {Ba}$_{1.2}${Mn}$_{8}${O}$_{16}$},}\ }\href@noop {} {\bibfield  {journal}
  {\bibinfo  {journal} {J. Phys.:Condens. Matt.}\ }\textbf {\bibinfo {volume}
  {18}},\ \bibinfo {pages} {3745} (\bibinfo {year} {2006})}\BibitemShut
  {NoStop}%
\bibitem [{\citenamefont {Luo}\ \emph {et~al.}(2010)\citenamefont {Luo},
  \citenamefont {Zhu}, \citenamefont {Liang}, \citenamefont {Rao},
  \citenamefont {Li},\ and\ \citenamefont {Du}}]{Luo10}%
  \BibitemOpen
  \bibfield  {author} {\bibinfo {author} {\bibfnamefont {J.}~\bibnamefont
  {Luo}}, \bibinfo {author} {\bibfnamefont {H.~T.}\ \bibnamefont {Zhu}},
  \bibinfo {author} {\bibfnamefont {J.~K.}\ \bibnamefont {Liang}}, \bibinfo
  {author} {\bibfnamefont {G.~H.}\ \bibnamefont {Rao}}, \bibinfo {author}
  {\bibfnamefont {J.~B.}\ \bibnamefont {Li}}, \ and\ \bibinfo {author}
  {\bibfnamefont {Z.~M.}\ \bibnamefont {Du}},\ }\bibfield  {title} {\enquote
  {\bibinfo {title} {Tuning magnetic properties of $\alpha$-{Mn}{O}$_{2}$
  nanotubes by {K}$^{+}$ doping},}\ }\href@noop {} {\bibfield  {journal}
  {\bibinfo  {journal} {J. Phys. Chem. C}\ }\textbf {\bibinfo {volume} {114}},\
  \bibinfo {pages} {8782--8786} (\bibinfo {year} {2010})}\BibitemShut {NoStop}%
\bibitem [{\citenamefont {Barudzija}\ \emph {et~al.}(2016)\citenamefont
  {Barudzija}, \citenamefont {Kusigerski}, \citenamefont {Cvjeticanin},
  \citenamefont {Sorgic}, \citenamefont {Perovic},\ and\ \citenamefont
  {Mitric}}]{Barudzija16}%
  \BibitemOpen
  \bibfield  {author} {\bibinfo {author} {\bibfnamefont {T.}~\bibnamefont
  {Barudzija}}, \bibinfo {author} {\bibfnamefont {V.}~\bibnamefont
  {Kusigerski}}, \bibinfo {author} {\bibfnamefont {N.}~\bibnamefont
  {Cvjeticanin}}, \bibinfo {author} {\bibfnamefont {S.}~\bibnamefont {Sorgic}},
  \bibinfo {author} {\bibfnamefont {M.}~\bibnamefont {Perovic}}, \ and\
  \bibinfo {author} {\bibfnamefont {M.}~\bibnamefont {Mitric}},\ }\bibfield
  {title} {\enquote {\bibinfo {title} {Structural and magnetic properties of
  hydrothermally synthesized $\beta$-{Mn}{O}$_{2}$ and
  $\alpha$-{K}$_{x}${Mn}{O}$_{2}$ nanorods},}\ }\href@noop {} {\bibfield
  {journal} {\bibinfo  {journal} {J. Alloy. Compd.}\ }\textbf {\bibinfo
  {volume} {665}},\ \bibinfo {pages} {261--270} (\bibinfo {year}
  {2016})}\BibitemShut {NoStop}%
\bibitem [{\citenamefont {Vicat}\ \emph {et~al.}(1986)\citenamefont {Vicat},
  \citenamefont {Fanchon}, \citenamefont {Strobel},\ and\ \citenamefont
  {Tran}}]{Vicat1986}%
  \BibitemOpen
  \bibfield  {author} {\bibinfo {author} {\bibfnamefont {J.}~\bibnamefont
  {Vicat}}, \bibinfo {author} {\bibfnamefont {E.}~\bibnamefont {Fanchon}},
  \bibinfo {author} {\bibfnamefont {P.}~\bibnamefont {Strobel}}, \ and\
  \bibinfo {author} {\bibfnamefont {Q.~D.}\ \bibnamefont {Tran}},\ }\bibfield
  {title} {\enquote {\bibinfo {title} {The structure of
  {K}$_{1.33}${Mn}$_{8}${O}$_{16}$ and cation ordering in hollandite-type
  structures},}\ }\href@noop {} {\bibfield  {journal} {\bibinfo  {journal}
  {Acta Crystallogr. Sect. B}\ }\textbf {\bibinfo {volume} {42}},\ \bibinfo
  {pages} {162--167} (\bibinfo {year} {1986})}\BibitemShut {NoStop}%
\bibitem [{\citenamefont {Kadoma}\ \emph {et~al.}(2007)\citenamefont {Kadoma},
  \citenamefont {Oshitari}, \citenamefont {Ui},\ and\ \citenamefont
  {Kumagai}}]{Kadoma2007}%
  \BibitemOpen
  \bibfield  {author} {\bibinfo {author} {\bibfnamefont {Y.}~\bibnamefont
  {Kadoma}}, \bibinfo {author} {\bibfnamefont {S.}~\bibnamefont {Oshitari}},
  \bibinfo {author} {\bibfnamefont {K.}~\bibnamefont {Ui}}, \ and\ \bibinfo
  {author} {\bibfnamefont {N.}~\bibnamefont {Kumagai}},\ }\bibfield  {title}
  {\enquote {\bibinfo {title} {Synthesis of hollandite-type {Li}$_x${Mn}{O}$_2$
  by {Li}$^+$ ion-exchange in molten salt and lithium insertion
  characteristics},}\ }\href@noop {} {\bibfield  {journal} {\bibinfo  {journal}
  {Electrochim. Acta}\ }\textbf {\bibinfo {volume} {53}},\ \bibinfo {pages}
  {1697 -- 1702} (\bibinfo {year} {2007})}\BibitemShut {NoStop}%
\bibitem [{\citenamefont {Chang}\ and\ \citenamefont
  {Jansen}(1984)}]{Jansen1984}%
  \BibitemOpen
  \bibfield  {author} {\bibinfo {author} {\bibfnamefont {F.~M.}\ \bibnamefont
  {Chang}}\ and\ \bibinfo {author} {\bibfnamefont {M.}~\bibnamefont {Jansen}},\
  }\bibfield  {title} {\enquote {\bibinfo {title}
  {{Ag}$_{1.8}${Mn}$_{8}${O}$_{16}$: Square planar coordinated {Ag}$^+$ ions in
  the channels of a novel hollandite variant},}\ }\href@noop {} {\bibfield
  {journal} {\bibinfo  {journal} {Angew. Chem. Int. Ed.}\ }\textbf {\bibinfo
  {volume} {23}},\ \bibinfo {pages} {906--907} (\bibinfo {year}
  {1984})}\BibitemShut {NoStop}%
\bibitem [{\citenamefont {Cockayne}\ and\ \citenamefont
  {Li}(2012)}]{COCKAYNE2012}%
  \BibitemOpen
  \bibfield  {author} {\bibinfo {author} {\bibfnamefont {E.}~\bibnamefont
  {Cockayne}}\ and\ \bibinfo {author} {\bibfnamefont {L.}~\bibnamefont {Li}},\
  }\bibfield  {title} {\enquote {\bibinfo {title} {First-principles {DFT+U}
  studies of the atomic, electronic, and magnetic structure of
  $\alpha$-{Mn}{O}$_{2}$ (cryptomelane)},}\ }\href@noop {} {\bibfield
  {journal} {\bibinfo  {journal} {Chem. Phys. Lett.}\ }\textbf {\bibinfo
  {volume} {544}},\ \bibinfo {pages} {53 -- 58} (\bibinfo {year}
  {2012})}\BibitemShut {NoStop}%
\bibitem [{\citenamefont {Crespo}\ and\ \citenamefont
  {Seriani}(2013)}]{Crespo13a}%
  \BibitemOpen
  \bibfield  {author} {\bibinfo {author} {\bibfnamefont {Y.}~\bibnamefont
  {Crespo}}\ and\ \bibinfo {author} {\bibfnamefont {N.}~\bibnamefont
  {Seriani}},\ }\bibfield  {title} {\enquote {\bibinfo {title} {Electronic and
  magnetic properties of $\ensuremath{\alpha}$-{Mn}{O}${}_{2}$ from ab initio
  calculations},}\ }\href@noop {} {\bibfield  {journal} {\bibinfo  {journal}
  {Phys. Rev. B}\ }\textbf {\bibinfo {volume} {88}},\ \bibinfo {pages} {144428}
  (\bibinfo {year} {2013})}\BibitemShut {NoStop}%
\bibitem [{\citenamefont {Ochoa}\ \emph {et~al.}(2016)\citenamefont {Ochoa},
  \citenamefont {Huang}, \citenamefont {Tang}, \citenamefont {Cocoletzi},\ and\
  \citenamefont {Springborg}}]{Ochoa16}%
  \BibitemOpen
  \bibfield  {author} {\bibinfo {author} {\bibfnamefont {F.~S.}\ \bibnamefont
  {Ochoa}}, \bibinfo {author} {\bibfnamefont {Z.}~\bibnamefont {Huang}},
  \bibinfo {author} {\bibfnamefont {X.}~\bibnamefont {Tang}}, \bibinfo {author}
  {\bibfnamefont {G.~H.}\ \bibnamefont {Cocoletzi}}, \ and\ \bibinfo {author}
  {\bibfnamefont {M.}~\bibnamefont {Springborg}},\ }\bibfield  {title}
  {\enquote {\bibinfo {title} {Magnetostructural phase transition assisted by
  temperature in {Ag}-$\ensuremath{\alpha}${Mn}{O}$_{2}$: {A} density
  functional theory study},}\ }\href@noop {} {\bibfield  {journal} {\bibinfo
  {journal} {Phys. Chem. Chem. Phys.}\ }\textbf {\bibinfo {volume} {18}},\
  \bibinfo {pages} {7442--7448} (\bibinfo {year} {2016})}\BibitemShut {NoStop}%
\bibitem [{\citenamefont {Kitchaev}\ \emph {et~al.}(2017)\citenamefont
  {Kitchaev}, \citenamefont {Dacek}, \citenamefont {Sun},\ and\ \citenamefont
  {Ceder}}]{Kitchaev17}%
  \BibitemOpen
  \bibfield  {author} {\bibinfo {author} {\bibfnamefont {D.~A.}\ \bibnamefont
  {Kitchaev}}, \bibinfo {author} {\bibfnamefont {S.~T.}\ \bibnamefont {Dacek}},
  \bibinfo {author} {\bibfnamefont {W.}~\bibnamefont {Sun}}, \ and\ \bibinfo
  {author} {\bibfnamefont {G.}~\bibnamefont {Ceder}},\ }\bibfield  {title}
  {\enquote {\bibinfo {title} {Thermodynamics of phase selection in
  {Mn}{O}$_{2}$ framework structures through alkali intercalation and
  hydration},}\ }\href@noop {} {\bibfield  {journal} {\bibinfo  {journal} {J.
  Am. Chem. Soc.}\ }\textbf {\bibinfo {volume} {139}},\ \bibinfo {pages}
  {2672--2681} (\bibinfo {year} {2017})}\BibitemShut {NoStop}%
\bibitem [{\citenamefont {Perdew}\ \emph {et~al.}(1996)\citenamefont {Perdew},
  \citenamefont {Burke},\ and\ \citenamefont {Ernzerhof}}]{Perdew96}%
  \BibitemOpen
  \bibfield  {author} {\bibinfo {author} {\bibfnamefont {J.~P.}\ \bibnamefont
  {Perdew}}, \bibinfo {author} {\bibfnamefont {K.}~\bibnamefont {Burke}}, \
  and\ \bibinfo {author} {\bibfnamefont {M.}~\bibnamefont {Ernzerhof}},\
  }\bibfield  {title} {\enquote {\bibinfo {title} {Generalized gradient
  approximation made simple},}\ }\href@noop {} {\bibfield  {journal} {\bibinfo
  {journal} {Phys. Rev. Lett.}\ }\textbf {\bibinfo {volume} {77}},\ \bibinfo
  {pages} {3865--3868} (\bibinfo {year} {1996})}\BibitemShut {NoStop}%
\bibitem [{\citenamefont {Klime\ifmmode\check{s}\else\v{s}\fi{}}\ \emph
  {et~al.}(2010)\citenamefont {Klime\ifmmode\check{s}\else\v{s}\fi{}},
  \citenamefont {Bowler},\ and\ \citenamefont {Michaelides}}]{OptB88}%
  \BibitemOpen
  \bibfield  {author} {\bibinfo {author} {\bibfnamefont {J.}~\bibnamefont
  {Klime\ifmmode\check{s}\else\v{s}\fi{}}}, \bibinfo {author} {\bibfnamefont
  {D.~R.}\ \bibnamefont {Bowler}}, \ and\ \bibinfo {author} {\bibfnamefont
  {A.}~\bibnamefont {Michaelides}},\ }\bibfield  {title} {\enquote {\bibinfo
  {title} {Chemical accuracy for the van der waals density functional},}\
  }\href@noop {} {\bibfield  {journal} {\bibinfo  {journal} {J. Phys.: Condens.
  Matt.}\ }\textbf {\bibinfo {volume} {22}},\ \bibinfo {pages} {022201}
  (\bibinfo {year} {2010})}\BibitemShut {NoStop}%
\bibitem [{\citenamefont {Anisimov}\ \emph {et~al.}(1991)\citenamefont
  {Anisimov}, \citenamefont {Zaanen},\ and\ \citenamefont
  {Andersen}}]{Anisimov91}%
  \BibitemOpen
  \bibfield  {author} {\bibinfo {author} {\bibfnamefont {V.~I.}\ \bibnamefont
  {Anisimov}}, \bibinfo {author} {\bibfnamefont {J.}~\bibnamefont {Zaanen}}, \
  and\ \bibinfo {author} {\bibfnamefont {O.~K.}\ \bibnamefont {Andersen}},\
  }\bibfield  {title} {\enquote {\bibinfo {title} {Band theory and {Mott}
  insulators: Hubbard {U} instead of {Stoner} {I}},}\ }\href@noop {} {\bibfield
   {journal} {\bibinfo  {journal} {Phys. Rev. B}\ }\textbf {\bibinfo {volume}
  {44}},\ \bibinfo {pages} {943--954} (\bibinfo {year} {1991})}\BibitemShut
  {NoStop}%
\bibitem [{\citenamefont {Anisimov}\ \emph {et~al.}(1997)\citenamefont
  {Anisimov}, \citenamefont {Aryasetiawan},\ and\ \citenamefont
  {Lichtenstein}}]{Anisimov97}%
  \BibitemOpen
  \bibfield  {author} {\bibinfo {author} {\bibfnamefont {V.~I.}\ \bibnamefont
  {Anisimov}}, \bibinfo {author} {\bibfnamefont {F.}~\bibnamefont
  {Aryasetiawan}}, \ and\ \bibinfo {author} {\bibfnamefont {A.~I.}\
  \bibnamefont {Lichtenstein}},\ }\bibfield  {title} {\enquote {\bibinfo
  {title} {First-principles calculations of the electronic structure and
  spectra of strongly correlated systems: {The} {LDA} + {U} method},}\
  }\href@noop {} {\bibfield  {journal} {\bibinfo  {journal} {J. Phys.: Condens.
  Matt.}\ }\textbf {\bibinfo {volume} {9}},\ \bibinfo {pages} {767} (\bibinfo
  {year} {1997})}\BibitemShut {NoStop}%
\bibitem [{\citenamefont {Dudarev}\ \emph {et~al.}(1998)\citenamefont
  {Dudarev}, \citenamefont {Botton}, \citenamefont {Savrasov}, \citenamefont
  {Humphreys},\ and\ \citenamefont {Sutton}}]{Dudarev98}%
  \BibitemOpen
  \bibfield  {author} {\bibinfo {author} {\bibfnamefont {S.~L.}\ \bibnamefont
  {Dudarev}}, \bibinfo {author} {\bibfnamefont {G.~A.}\ \bibnamefont {Botton}},
  \bibinfo {author} {\bibfnamefont {S.~Y.}\ \bibnamefont {Savrasov}}, \bibinfo
  {author} {\bibfnamefont {C.~J.}\ \bibnamefont {Humphreys}}, \ and\ \bibinfo
  {author} {\bibfnamefont {A.~P.}\ \bibnamefont {Sutton}},\ }\bibfield  {title}
  {\enquote {\bibinfo {title} {Electron-energy-loss spectra and the structural
  stability of nickel oxide: An {LSDA+U} study},}\ }\href@noop {} {\bibfield
  {journal} {\bibinfo  {journal} {Phys. Rev. B}\ }\textbf {\bibinfo {volume}
  {57}},\ \bibinfo {pages} {1505--1509} (\bibinfo {year} {1998})}\BibitemShut
  {NoStop}%
\bibitem [{\citenamefont {Liechtenstein}\ \emph {et~al.}(1995)\citenamefont
  {Liechtenstein}, \citenamefont {Anisimov},\ and\ \citenamefont
  {Zaanen}}]{Liechtenstein95}%
  \BibitemOpen
  \bibfield  {author} {\bibinfo {author} {\bibfnamefont {A.~I.}\ \bibnamefont
  {Liechtenstein}}, \bibinfo {author} {\bibfnamefont {V.~I.}\ \bibnamefont
  {Anisimov}}, \ and\ \bibinfo {author} {\bibfnamefont {J.}~\bibnamefont
  {Zaanen}},\ }\bibfield  {title} {\enquote {\bibinfo {title}
  {Density-functional theory and strong interactions: Orbital ordering in
  {Mott}-{Hubbard} insulators},}\ }\href@noop {} {\bibfield  {journal}
  {\bibinfo  {journal} {Phys. Rev. B}\ }\textbf {\bibinfo {volume} {52}},\
  \bibinfo {pages} {5467--5470} (\bibinfo {year} {1995})}\BibitemShut {NoStop}%
\bibitem [{\citenamefont {Heyd}\ \emph {et~al.}(2003)\citenamefont {Heyd},
  \citenamefont {Scuseria},\ and\ \citenamefont {Ernzerhof}}]{HSE03}%
  \BibitemOpen
  \bibfield  {author} {\bibinfo {author} {\bibfnamefont {J.}~\bibnamefont
  {Heyd}}, \bibinfo {author} {\bibfnamefont {G.~E.}\ \bibnamefont {Scuseria}},
  \ and\ \bibinfo {author} {\bibfnamefont {M.}~\bibnamefont {Ernzerhof}},\
  }\bibfield  {title} {\enquote {\bibinfo {title} {Hybrid functionals based on
  a screened {Coulomb} potential},}\ }\href@noop {} {\bibfield  {journal}
  {\bibinfo  {journal} {J. Chem. Phys.}\ }\textbf {\bibinfo {volume} {118}},\
  \bibinfo {pages} {8207--8215} (\bibinfo {year} {2003})}\BibitemShut {NoStop}%
\bibitem [{\citenamefont {Heyd}\ \emph {et~al.}(2006)\citenamefont {Heyd},
  \citenamefont {Scuseria},\ and\ \citenamefont {Ernzerhof}}]{HSE06}%
  \BibitemOpen
  \bibfield  {author} {\bibinfo {author} {\bibfnamefont {Jochen}\ \bibnamefont
  {Heyd}}, \bibinfo {author} {\bibfnamefont {Gustavo~E.}\ \bibnamefont
  {Scuseria}}, \ and\ \bibinfo {author} {\bibfnamefont {Matthias}\ \bibnamefont
  {Ernzerhof}},\ }\bibfield  {title} {\enquote {\bibinfo {title} {Erratum:
  "{Hybrid} functionals based on a screened {Coulomb} potential" [{J.} {Chem.}
  {Phys.} 118, 8207 (2003)]},}\ }\href@noop {} {\bibfield  {journal} {\bibinfo
  {journal} {J. Chem. Phys.}\ }\textbf {\bibinfo {volume} {124}},\ \bibinfo
  {pages} {219906} (\bibinfo {year} {2006})}\BibitemShut {NoStop}%
\bibitem [{\citenamefont {Franchini}(2014)}]{Franchini14}%
  \BibitemOpen
  \bibfield  {author} {\bibinfo {author} {\bibfnamefont {C.}~\bibnamefont
  {Franchini}},\ }\bibfield  {title} {\enquote {\bibinfo {title} {Hybrid
  functionals applied to perovskites},}\ }\href@noop {} {\bibfield  {journal}
  {\bibinfo  {journal} {J. Phys.: Condens. Matt.}\ }\textbf {\bibinfo {volume}
  {26}} (\bibinfo {year} {2014})}\BibitemShut {NoStop}%
\bibitem [{\citenamefont {Sun}\ \emph {et~al.}(2015)\citenamefont {Sun},
  \citenamefont {Ruzsinszky},\ and\ \citenamefont {Perdew}}]{SCAN}%
  \BibitemOpen
  \bibfield  {author} {\bibinfo {author} {\bibfnamefont {J.}~\bibnamefont
  {Sun}}, \bibinfo {author} {\bibfnamefont {A.}~\bibnamefont {Ruzsinszky}}, \
  and\ \bibinfo {author} {\bibfnamefont {J.~P.}\ \bibnamefont {Perdew}},\
  }\bibfield  {title} {\enquote {\bibinfo {title} {Strongly constrained and
  appropriately normed semilocal density functional},}\ }\href@noop {}
  {\bibfield  {journal} {\bibinfo  {journal} {Phys. Rev. Lett.}\ }\textbf
  {\bibinfo {volume} {115}},\ \bibinfo {pages} {036402} (\bibinfo {year}
  {2015})}\BibitemShut {NoStop}%
\bibitem [{\citenamefont {Franchini}\ \emph {et~al.}(2007)\citenamefont
  {Franchini}, \citenamefont {Podloucky}, \citenamefont {Paier}, \citenamefont
  {Marsman},\ and\ \citenamefont {Kresse}}]{Franchini07}%
  \BibitemOpen
  \bibfield  {author} {\bibinfo {author} {\bibfnamefont {C.}~\bibnamefont
  {Franchini}}, \bibinfo {author} {\bibfnamefont {R.}~\bibnamefont
  {Podloucky}}, \bibinfo {author} {\bibfnamefont {J.}~\bibnamefont {Paier}},
  \bibinfo {author} {\bibfnamefont {M.}~\bibnamefont {Marsman}}, \ and\
  \bibinfo {author} {\bibfnamefont {G.}~\bibnamefont {Kresse}},\ }\bibfield
  {title} {\enquote {\bibinfo {title} {Ground-state properties of multivalent
  manganese oxides: Density functional and hybrid density functional
  calculations},}\ }\href@noop {} {\bibfield  {journal} {\bibinfo  {journal}
  {Phys. Rev. B}\ }\textbf {\bibinfo {volume} {75}},\ \bibinfo {pages} {195128}
  (\bibinfo {year} {2007})}\BibitemShut {NoStop}%
\bibitem [{\citenamefont {Tompsett}\ \emph {et~al.}(2012)\citenamefont
  {Tompsett}, \citenamefont {Middlemiss},\ and\ \citenamefont
  {Islam}}]{Tompsett12}%
  \BibitemOpen
  \bibfield  {author} {\bibinfo {author} {\bibfnamefont {D.~A.}\ \bibnamefont
  {Tompsett}}, \bibinfo {author} {\bibfnamefont {D.~S.}\ \bibnamefont
  {Middlemiss}}, \ and\ \bibinfo {author} {\bibfnamefont {M.~S.}\ \bibnamefont
  {Islam}},\ }\bibfield  {title} {\enquote {\bibinfo {title} {Importance of
  anisotropic coulomb interactions and exchange to the band gap and
  antiferromagnetism of $\ensuremath{\beta}$-{Mn}{O}${}_{2}$ from {DFT+U}},}\
  }\href@noop {} {\bibfield  {journal} {\bibinfo  {journal} {Phys. Rev. B}\
  }\textbf {\bibinfo {volume} {86}},\ \bibinfo {pages} {205126} (\bibinfo
  {year} {2012})}\BibitemShut {NoStop}%
\bibitem [{\citenamefont {Lim}\ \emph {et~al.}(2016)\citenamefont {Lim},
  \citenamefont {Saldana-Greco},\ and\ \citenamefont {Rappe}}]{Lim16}%
  \BibitemOpen
  \bibfield  {author} {\bibinfo {author} {\bibfnamefont {J.~S.}\ \bibnamefont
  {Lim}}, \bibinfo {author} {\bibfnamefont {D.}~\bibnamefont {Saldana-Greco}},
  \ and\ \bibinfo {author} {\bibfnamefont {A.~M.}\ \bibnamefont {Rappe}},\
  }\bibfield  {title} {\enquote {\bibinfo {title} {Improved pseudopotential
  transferability for magnetic and electronic properties of binary manganese
  oxides from {DFT}+{U}+{J} calculations},}\ }\href@noop {} {\bibfield
  {journal} {\bibinfo  {journal} {Phys. Rev. B}\ }\textbf {\bibinfo {volume}
  {94}},\ \bibinfo {pages} {165151} (\bibinfo {year} {2016})}\BibitemShut
  {NoStop}%
\bibitem [{\citenamefont {Kitchaev}\ \emph {et~al.}(2016)\citenamefont
  {Kitchaev}, \citenamefont {Peng}, \citenamefont {Liu}, \citenamefont {Sun},
  \citenamefont {Perdew},\ and\ \citenamefont {Ceder}}]{Kitchaev16}%
  \BibitemOpen
  \bibfield  {author} {\bibinfo {author} {\bibfnamefont {D.~A.}\ \bibnamefont
  {Kitchaev}}, \bibinfo {author} {\bibfnamefont {H.}~\bibnamefont {Peng}},
  \bibinfo {author} {\bibfnamefont {Y.}~\bibnamefont {Liu}}, \bibinfo {author}
  {\bibfnamefont {J.}~\bibnamefont {Sun}}, \bibinfo {author} {\bibfnamefont
  {J.~P.}\ \bibnamefont {Perdew}}, \ and\ \bibinfo {author} {\bibfnamefont
  {G.}~\bibnamefont {Ceder}},\ }\bibfield  {title} {\enquote {\bibinfo {title}
  {Energetics of {Mn}{O}$_{2}$ polymorphs in density functional theory},}\
  }\href@noop {} {\bibfield  {journal} {\bibinfo  {journal} {Phys. Rev. B}\
  }\textbf {\bibinfo {volume} {93}},\ \bibinfo {pages} {045132} (\bibinfo
  {year} {2016})}\BibitemShut {NoStop}%
\bibitem [{\citenamefont {Li}\ \emph {et~al.}(2013)\citenamefont {Li},
  \citenamefont {Cockayne}, \citenamefont {Williamson}, \citenamefont
  {Espinal},\ and\ \citenamefont {Wong-Ng}}]{LI2013120}%
  \BibitemOpen
  \bibfield  {author} {\bibinfo {author} {\bibfnamefont {L.}~\bibnamefont
  {Li}}, \bibinfo {author} {\bibfnamefont {E.}~\bibnamefont {Cockayne}},
  \bibinfo {author} {\bibfnamefont {I.}~\bibnamefont {Williamson}}, \bibinfo
  {author} {\bibfnamefont {L.}~\bibnamefont {Espinal}}, \ and\ \bibinfo
  {author} {\bibfnamefont {W.}~\bibnamefont {Wong-Ng}},\ }\bibfield  {title}
  {\enquote {\bibinfo {title} {First-principles studies of carbon dioxide
  adsorption in cryptomelane/hollandite-type manganese dioxide},}\ }\href@noop
  {} {\bibfield  {journal} {\bibinfo  {journal} {Chem. Phys. Lett.}\ }\textbf
  {\bibinfo {volume} {580}},\ \bibinfo {pages} {120 -- 125} (\bibinfo {year}
  {2013})}\BibitemShut {NoStop}%
\bibitem [{\citenamefont {Crespo}\ \emph {et~al.}(2013)\citenamefont {Crespo},
  \citenamefont {Andreanov},\ and\ \citenamefont {Seriani}}]{Crespo13b}%
  \BibitemOpen
  \bibfield  {author} {\bibinfo {author} {\bibfnamefont {Y.}~\bibnamefont
  {Crespo}}, \bibinfo {author} {\bibfnamefont {A.}~\bibnamefont {Andreanov}}, \
  and\ \bibinfo {author} {\bibfnamefont {N.}~\bibnamefont {Seriani}},\
  }\bibfield  {title} {\enquote {\bibinfo {title} {Competing antiferromagnetic
  and spin-glass phases in a hollandite structure},}\ }\href@noop {} {\bibfield
   {journal} {\bibinfo  {journal} {Phys. Rev. B}\ }\textbf {\bibinfo {volume}
  {88}},\ \bibinfo {pages} {014202} (\bibinfo {year} {2013})}\BibitemShut
  {NoStop}%
\bibitem [{\citenamefont {Jahn}\ and\ \citenamefont
  {Teller}(1937)}]{JahnTeller1937}%
  \BibitemOpen
  \bibfield  {author} {\bibinfo {author} {\bibfnamefont {H.~A.}\ \bibnamefont
  {Jahn}}\ and\ \bibinfo {author} {\bibfnamefont {E.}~\bibnamefont {Teller}},\
  }\bibfield  {title} {\enquote {\bibinfo {title} {Stability of polyatomic
  molecules in degenerate electronic states. {I.} {Orbital} degeneracy},}\
  }\href@noop {} {\bibfield  {journal} {\bibinfo  {journal} {Proc. Roy. Soc.
  A}\ }\textbf {\bibinfo {volume} {161}},\ \bibinfo {pages} {220--235}
  (\bibinfo {year} {1937})}\BibitemShut {NoStop}%
\bibitem [{\citenamefont {Vleck}(1939)}]{VanVleck1939}%
  \BibitemOpen
  \bibfield  {author} {\bibinfo {author} {\bibfnamefont {J.~H.~Van}\
  \bibnamefont {Vleck}},\ }\bibfield  {title} {\enquote {\bibinfo {title} {The
  {Jahn}-{Teller} effect and crystalline stark splitting for clusters of the
  form {X}{Y}$_{6}$},}\ }\href@noop {} {\bibfield  {journal} {\bibinfo
  {journal} {J. Chem. Phys.}\ }\textbf {\bibinfo {volume} {7}},\ \bibinfo
  {pages} {72--84} (\bibinfo {year} {1939})}\BibitemShut {NoStop}%
\bibitem [{\citenamefont {Kresse}\ and\ \citenamefont
  {Joubert}(1999)}]{Kresse99}%
  \BibitemOpen
  \bibfield  {author} {\bibinfo {author} {\bibfnamefont {G.}~\bibnamefont
  {Kresse}}\ and\ \bibinfo {author} {\bibfnamefont {D.}~\bibnamefont
  {Joubert}},\ }\bibfield  {title} {\enquote {\bibinfo {title} {From ultrasoft
  pseudopotentials to the projector augmented-wave method},}\ }\href@noop {}
  {\bibfield  {journal} {\bibinfo  {journal} {Phys. Rev. B}\ }\textbf {\bibinfo
  {volume} {59}},\ \bibinfo {pages} {1758--1775} (\bibinfo {year}
  {1999})}\BibitemShut {NoStop}%
\bibitem [{\citenamefont {Bl\"ochl}(1994)}]{Blochl94}%
  \BibitemOpen
  \bibfield  {author} {\bibinfo {author} {\bibfnamefont {P.~E.}\ \bibnamefont
  {Bl\"ochl}},\ }\bibfield  {title} {\enquote {\bibinfo {title} {Projector
  augmented-wave method},}\ }\href@noop {} {\bibfield  {journal} {\bibinfo
  {journal} {Phys. Rev. B}\ }\textbf {\bibinfo {volume} {50}},\ \bibinfo
  {pages} {17953--17979} (\bibinfo {year} {1994})}\BibitemShut {NoStop}%
\bibitem [{\citenamefont {Marini}\ \emph {et~al.}(2001)\citenamefont {Marini},
  \citenamefont {Onida},\ and\ \citenamefont {Del~Sole}}]{Marini01}%
  \BibitemOpen
  \bibfield  {author} {\bibinfo {author} {\bibfnamefont {A.}~\bibnamefont
  {Marini}}, \bibinfo {author} {\bibfnamefont {G.}~\bibnamefont {Onida}}, \
  and\ \bibinfo {author} {\bibfnamefont {R.}~\bibnamefont {Del~Sole}},\
  }\bibfield  {title} {\enquote {\bibinfo {title} {Quasiparticle electronic
  structure of copper in the $\mathit{GW}$ approximation},}\ }\href@noop {}
  {\bibfield  {journal} {\bibinfo  {journal} {Phys. Rev. Lett.}\ }\textbf
  {\bibinfo {volume} {88}},\ \bibinfo {pages} {016403} (\bibinfo {year}
  {2001})}\BibitemShut {NoStop}%
\bibitem [{\citenamefont {Paier}\ \emph {et~al.}(2006)\citenamefont {Paier},
  \citenamefont {Marsman}, \citenamefont {Hummer}, \citenamefont {Kresse},
  \citenamefont {Gerber},\ and\ \citenamefont {Angyan}}]{Paier06}%
  \BibitemOpen
  \bibfield  {author} {\bibinfo {author} {\bibfnamefont {J.}~\bibnamefont
  {Paier}}, \bibinfo {author} {\bibfnamefont {M.}~\bibnamefont {Marsman}},
  \bibinfo {author} {\bibfnamefont {K.}~\bibnamefont {Hummer}}, \bibinfo
  {author} {\bibfnamefont {G.}~\bibnamefont {Kresse}}, \bibinfo {author}
  {\bibfnamefont {I.~C.}\ \bibnamefont {Gerber}}, \ and\ \bibinfo {author}
  {\bibfnamefont {J.~G.}\ \bibnamefont {Angyan}},\ }\bibfield  {title}
  {\enquote {\bibinfo {title} {Screened hybrid density functionals applied to
  solids},}\ }\href@noop {} {\bibfield  {journal} {\bibinfo  {journal} {J.
  Chem. Phys.}\ }\textbf {\bibinfo {volume} {124}},\ \bibinfo {pages} {154709}
  (\bibinfo {year} {2006})}\BibitemShut {NoStop}%
\bibitem [{\citenamefont {Engel}(2009)}]{Engel09}%
  \BibitemOpen
  \bibfield  {author} {\bibinfo {author} {\bibfnamefont {E.}~\bibnamefont
  {Engel}},\ }\bibfield  {title} {\enquote {\bibinfo {title} {Relevance of
  core-valence interaction for electronic structure calculations with exact
  exchange},}\ }\href@noop {} {\bibfield  {journal} {\bibinfo  {journal} {Phys.
  Rev. B}\ }\textbf {\bibinfo {volume} {80}},\ \bibinfo {pages} {161205}
  (\bibinfo {year} {2009})}\BibitemShut {NoStop}%
\bibitem [{\citenamefont {Marzari}\ and\ \citenamefont
  {Vanderbilt}(1997)}]{Marzari97}%
  \BibitemOpen
  \bibfield  {author} {\bibinfo {author} {\bibfnamefont {N.}~\bibnamefont
  {Marzari}}\ and\ \bibinfo {author} {\bibfnamefont {D.}~\bibnamefont
  {Vanderbilt}},\ }\bibfield  {title} {\enquote {\bibinfo {title} {Maximally
  localized generalized wannier functions for composite energy bands},}\
  }\href@noop {} {\bibfield  {journal} {\bibinfo  {journal} {Phys. Rev. B}\
  }\textbf {\bibinfo {volume} {56}},\ \bibinfo {pages} {12847--12865} (\bibinfo
  {year} {1997})}\BibitemShut {NoStop}%
\bibitem [{\citenamefont {Mostofi}\ \emph {et~al.}(2008)\citenamefont
  {Mostofi}, \citenamefont {Yates}, \citenamefont {Lee}, \citenamefont {Souza},
  \citenamefont {Vanderbilt},\ and\ \citenamefont {Marzari}}]{wannier90}%
  \BibitemOpen
  \bibfield  {author} {\bibinfo {author} {\bibfnamefont {A.~A.}\ \bibnamefont
  {Mostofi}}, \bibinfo {author} {\bibfnamefont {J.~R.}\ \bibnamefont {Yates}},
  \bibinfo {author} {\bibfnamefont {Y.-S.}\ \bibnamefont {Lee}}, \bibinfo
  {author} {\bibfnamefont {I.}~\bibnamefont {Souza}}, \bibinfo {author}
  {\bibfnamefont {D.}~\bibnamefont {Vanderbilt}}, \ and\ \bibinfo {author}
  {\bibfnamefont {N.}~\bibnamefont {Marzari}},\ }\bibfield  {title} {\enquote
  {\bibinfo {title} {wannier90: A tool for obtaining maximally-localised
  wannier functions},}\ }\href@noop {} {\bibfield  {journal} {\bibinfo
  {journal} {Comp. Phys. Commun.}\ }\textbf {\bibinfo {volume} {178}},\
  \bibinfo {pages} {685 -- 699} (\bibinfo {year} {2008})}\BibitemShut {NoStop}%
\bibitem [{\citenamefont {Kramida}\ \emph {et~al.}(2015)\citenamefont
  {Kramida}, \citenamefont {{Yu.~Ralchenko}}, \citenamefont {Reader},\ and\
  \citenamefont {{NIST ASD Team}}}]{NIST_ASD}%
  \BibitemOpen
  \bibfield  {author} {\bibinfo {author} {\bibfnamefont {A.}~\bibnamefont
  {Kramida}}, \bibinfo {author} {\bibnamefont {{Yu.~Ralchenko}}}, \bibinfo
  {author} {\bibfnamefont {J.}~\bibnamefont {Reader}}, \ and\ \bibinfo {author}
  {\bibnamefont {{NIST ASD Team}}},\ }\href@noop {} {}\bibinfo {howpublished}
  {{NIST Atomic Spectra Database (ver. 5.3), [Online, 2017, June 24]. National
  Institute of Standards and Technology }} (\bibinfo {year} {2015})\BibitemShut
  {NoStop}%
\bibitem [{\citenamefont {Grau-Crespo}\ \emph {et~al.}(2007)\citenamefont
  {Grau-Crespo}, \citenamefont {Hamad}, \citenamefont {Catlow},\ and\
  \citenamefont {de~Leeuw}}]{SOD2007}%
  \BibitemOpen
  \bibfield  {author} {\bibinfo {author} {\bibfnamefont {R.}~\bibnamefont
  {Grau-Crespo}}, \bibinfo {author} {\bibfnamefont {S.}~\bibnamefont {Hamad}},
  \bibinfo {author} {\bibfnamefont {C.~R.~A.}\ \bibnamefont {Catlow}}, \ and\
  \bibinfo {author} {\bibfnamefont {N.~H.}\ \bibnamefont {de~Leeuw}},\
  }\bibfield  {title} {\enquote {\bibinfo {title} {Symmetry-adapted
  configurational modelling of fractional site occupancy in solids},}\
  }\href@noop {} {\bibfield  {journal} {\bibinfo  {journal} {J. Phys.: Condens.
  Matt.}\ }\textbf {\bibinfo {volume} {19}},\ \bibinfo {pages} {256201}
  (\bibinfo {year} {2007})}\BibitemShut {NoStop}%
\bibitem [{Note1()}]{Note1}%
  \BibitemOpen
  \bibinfo {note} {For example, we found 79 symmetrically distinct ways to
  distribute two Mn$^{3+}$ in the $1\times 1\times 2$ super cell and more than
  2900 configurations for a $1\times 1\times 3$ super cell.}\BibitemShut
  {Stop}%
\bibitem [{Note2()}]{Note2}%
  \BibitemOpen
  \bibinfo {note} {The remaining two charge orderings for the unit cell are not
  stable under ionic force minimization for both AF and FM spin
  alignment.}\BibitemShut {Stop}%
\bibitem [{\citenamefont {Fukuzawa}\ \emph {et~al.}(2013)\citenamefont
  {Fukuzawa}, \citenamefont {Ootsuki},\ and\ \citenamefont
  {Mizokawa}}]{Fukuzawa13}%
  \BibitemOpen
  \bibfield  {author} {\bibinfo {author} {\bibfnamefont {M.}~\bibnamefont
  {Fukuzawa}}, \bibinfo {author} {\bibfnamefont {D.}~\bibnamefont {Ootsuki}}, \
  and\ \bibinfo {author} {\bibfnamefont {T.}~\bibnamefont {Mizokawa}},\
  }\bibfield  {title} {\enquote {\bibinfo {title} {Spin-charge-orbital ordering
  in hollandite-type manganites studied by model {Hartree-Fock} calculation},}\
  }\href@noop {} {\bibfield  {journal} {\bibinfo  {journal} {J. Phys. Soc.
  Jpn.}\ }\textbf {\bibinfo {volume} {82}},\ \bibinfo {pages} {074708}
  (\bibinfo {year} {2013})}\BibitemShut {NoStop}%
\bibitem [{\citenamefont {Rodriguez-Carvajal}\ \emph
  {et~al.}(1998)\citenamefont {Rodriguez-Carvajal}, \citenamefont {Rousse},
  \citenamefont {Masquelier},\ and\ \citenamefont
  {Hervieu}}]{RodriguezCarvajal98}%
  \BibitemOpen
  \bibfield  {author} {\bibinfo {author} {\bibfnamefont {J.}~\bibnamefont
  {Rodriguez-Carvajal}}, \bibinfo {author} {\bibfnamefont {G.}~\bibnamefont
  {Rousse}}, \bibinfo {author} {\bibfnamefont {C.}~\bibnamefont {Masquelier}},
  \ and\ \bibinfo {author} {\bibfnamefont {M.}~\bibnamefont {Hervieu}},\
  }\bibfield  {title} {\enquote {\bibinfo {title} {Electronic crystallization
  in a lithium battery material: Columnar ordering of electrons and holes in
  the spinel {Li}{Mn}$_{2}${O}$_{4}$},}\ }\href@noop {} {\bibfield  {journal}
  {\bibinfo  {journal} {Phys. Rev. Lett.}\ }\textbf {\bibinfo {volume} {81}},\
  \bibinfo {pages} {4660--4663} (\bibinfo {year} {1998})}\BibitemShut {NoStop}%
\bibitem [{\citenamefont {Massarotti}\ \emph {et~al.}(1999)\citenamefont
  {Massarotti}, \citenamefont {Capsoni}, \citenamefont {Bini}, \citenamefont
  {Scardi}, \citenamefont {Leoni}, \citenamefont {Baron},\ and\ \citenamefont
  {Berg}}]{Massarotti99}%
  \BibitemOpen
  \bibfield  {author} {\bibinfo {author} {\bibfnamefont {V.}~\bibnamefont
  {Massarotti}}, \bibinfo {author} {\bibfnamefont {D.}~\bibnamefont {Capsoni}},
  \bibinfo {author} {\bibfnamefont {M.}~\bibnamefont {Bini}}, \bibinfo {author}
  {\bibfnamefont {P.}~\bibnamefont {Scardi}}, \bibinfo {author} {\bibfnamefont
  {M.}~\bibnamefont {Leoni}}, \bibinfo {author} {\bibfnamefont
  {V.}~\bibnamefont {Baron}}, \ and\ \bibinfo {author} {\bibfnamefont
  {H.}~\bibnamefont {Berg}},\ }\bibfield  {title} {\enquote {\bibinfo {title}
  {{Li}{Mn}$_{2}${O}$_{4}$ low-temperature phase: synchrotron and neutron
  diffraction study},}\ }\href@noop {} {\bibfield  {journal} {\bibinfo
  {journal} {J. Appl. Cryst.}\ }\textbf {\bibinfo {volume} {32}},\ \bibinfo
  {pages} {1186--1189} (\bibinfo {year} {1999})}\BibitemShut {NoStop}%
\bibitem [{\citenamefont {Ouyang}\ \emph {et~al.}(2009)\citenamefont {Ouyang},
  \citenamefont {Shi},\ and\ \citenamefont {Lei}}]{Ouyang09}%
  \BibitemOpen
  \bibfield  {author} {\bibinfo {author} {\bibfnamefont {C.~Y.}\ \bibnamefont
  {Ouyang}}, \bibinfo {author} {\bibfnamefont {S.~Q.}\ \bibnamefont {Shi}}, \
  and\ \bibinfo {author} {\bibfnamefont {M.~S.}\ \bibnamefont {Lei}},\
  }\bibfield  {title} {\enquote {\bibinfo {title} {{Jahn}-{Teller} distortion
  and electronic structure of {Li}{Mn}$_{2}${O}$_{4}$},}\ }\href@noop {}
  {\bibfield  {journal} {\bibinfo  {journal} {J. Alloy. Compd.}\ }\textbf
  {\bibinfo {volume} {474}},\ \bibinfo {pages} {370--374} (\bibinfo {year}
  {2009})}\BibitemShut {NoStop}%
\bibitem [{\citenamefont {Shannon}(1976)}]{Shannon76}%
  \BibitemOpen
  \bibfield  {author} {\bibinfo {author} {\bibfnamefont {R.~D.}\ \bibnamefont
  {Shannon}},\ }\bibfield  {title} {\enquote {\bibinfo {title} {{Revised
  effective ionic radii and systematic studies of interatomic distances in
  halides and chalcogenides}},}\ }\href@noop {} {\bibfield  {journal} {\bibinfo
   {journal} {Acta Crystallogr. Sect. A}\ }\textbf {\bibinfo {volume} {32}},\
  \bibinfo {pages} {751--767} (\bibinfo {year} {1976})}\BibitemShut {NoStop}%
\bibitem [{\citenamefont {Anderson}(1950)}]{Anderson50}%
  \BibitemOpen
  \bibfield  {author} {\bibinfo {author} {\bibfnamefont {P.~W.}\ \bibnamefont
  {Anderson}},\ }\bibfield  {title} {\enquote {\bibinfo {title}
  {Antiferromagnetism. {Theory} of superexchange interaction},}\ }\href@noop {}
  {\bibfield  {journal} {\bibinfo  {journal} {Phys. Rev.}\ }\textbf {\bibinfo
  {volume} {79}},\ \bibinfo {pages} {350--356} (\bibinfo {year}
  {1950})}\BibitemShut {NoStop}%
\bibitem [{\citenamefont {Goodenough}(1955)}]{Goodenough55}%
  \BibitemOpen
  \bibfield  {author} {\bibinfo {author} {\bibfnamefont {J.~B.}\ \bibnamefont
  {Goodenough}},\ }\bibfield  {title} {\enquote {\bibinfo {title} {Theory of
  the role of covalence in the perovskite-type manganites [{La},
  {M}({II})]{Mn}{O}$_{3}$},}\ }\href@noop {} {\bibfield  {journal} {\bibinfo
  {journal} {Phys. Rev.}\ }\textbf {\bibinfo {volume} {100}},\ \bibinfo {pages}
  {564--573} (\bibinfo {year} {1955})}\BibitemShut {NoStop}%
\bibitem [{\citenamefont {Kanamori}(1960)}]{Kanamori1960}%
  \BibitemOpen
  \bibfield  {author} {\bibinfo {author} {\bibfnamefont {J.}~\bibnamefont
  {Kanamori}},\ }\bibfield  {title} {\enquote {\bibinfo {title} {Crystal
  distortion in magnetic compounds},}\ }\href@noop {} {\bibfield  {journal}
  {\bibinfo  {journal} {J. Appl. Phys.}\ }\textbf {\bibinfo {volume} {31}},\
  \bibinfo {pages} {S14--S23} (\bibinfo {year} {1960})}\BibitemShut {NoStop}%
\end{thebibliography}%

\end{document}